\documentstyle[epsfig]{article}

\begin{document}
\title{Time-of-arrival probabilities and quantum measurements}
\author{Charis Anastopoulos\footnote{anastop@physics.upatras.gr}\\
{\small Department of Physics, University of Patras, 26500 Patras,
Greece}
\\
 and \\ Ntina Savvidou \footnote{ntina@imperial.ac.uk} \\
 {\small  Theoretical Physics Group, Imperial College, SW7 2BZ,
London, UK}}

\maketitle

\begin{abstract}
We study the construction of probability densities for
time-of-arrival in quantum mechanics. Our treatment is based upon
the facts that (i) time appears in quantum theory as an external
parameter to the system, and (ii) propositions about the
time-of-arrival appear naturally when one considers histories. The
definition of time-of-arrival probabilities is  straightforward in
stochastic processes. The difficulties that arise in quantum theory
are due to the fact that the time parameter of Schr\"odinger's
equation does not naturally define a probability density at the
continuum limit, but also because the procedure one follows is
sensitive on the interpretation of the reduction procedure. We
consider the issue in Copenhagen quantum mechanics and in
history-based schemes like consistent histories. The benefit of the
latter is that it allows a proper passage to the continuous
limit--there are however problems related to the quantum Zeno effect
and decoherence. We finally employ the histories-based description
to construct Positive-Operator-Valued-Measures (POVMs) for the
time-of-arrival, which are valid for a general Hamiltonian. These
POVMs typically depend on the resolution of the measurement device;
for a free particle, however,  this dependence cancels in the
physically relevant regime and the POVM coincides with that of
Kijowski.

\end{abstract}

\renewcommand {\thesection}{\arabic{section}}
\renewcommand {\theequation}{\thesection. \arabic{equation}}
\let \ssection = \section \renewcommand{\section}{\setcounter{equation}{0} \ssection}

\section{Introduction}
The probabilities provided by quantum mechanics usually refer to
measurements that take place at specific moments of time. However,
there are physically realisable measurements that do not fall in
this category. One such example involves the determination of a
particle's time-of-arrival (or time-of-flight) measurement. If the
quantum mechanical system is described by a wave function
$\psi(x,t)$, its modulus square $|\psi(x,t)|^2$ is a probability
density with respect to $x$ at any time $t$. If $\psi$ describes a
particle beam, the probability density  describes the particles'
distribution in space, as if a snapshot were taken at a moment $t$.
However, the set-up for particle detection is slightly different.
The time interval between the emission of the beam and its detection
is not fixed; rather one places a detector at a fixed distance $L$
from the source. This is {\em not} a single-time measurement of the
particle's position. Therefore, the object $|\psi(x,t)|^2$ is not
immediately relevant, because it is not a probability distribution
with respect to $t$. Only if one assumes that the initial
wave-packet is narrowly concentrated around a specific momentum
value $p$ and that it remains so at all times until detection, is it
possible to state that this measurement is equivalent to a
single-time measurement at time $t = m L/ p$ ($m$ being the particle
mass). But even for small momentum spreads of the initial state, the
assumptions above are valid only for free particles.

 In the general case, the detector  registers
particles at different times $t$. One therefore needs to construct a
probability $p(t)$ for the time-of-arrival.  This would have been an
elementary problem, if there existed an operator representing time
in the system's Hilbert space.  In that case one would use the Born
rule to determine a probability density for the time-of-arrival.
Unfortunately the existence of a time operator $\hat{T}$, conjugate
to the Hamiltonian (so that the Hamiltonian generates
time-translations: $e^{i\hat{H}s} \hat{T} e^{-i \hat{H}s} = \hat{T}
+ s$) is ruled out by the requirement that the Hamiltonian is
bounded from below \cite{PaEn, UnWa89}. Nonetheless, one may still
define a quantum time variable, by choosing some degrees of freedom
of the system (or of a measuring apparatus) as defining an internal
"clock". The simplest example for the time-of-arrival is the quantum
version of the classical function $mx/p$, for a free particle of
mass $m$--$x, p$ being the position and momentum respectively.
Still, clock times fail to be conjugate to the Hamiltonian of the
system, with the result that they do not forward under Hamiltonian
evolution. In simpler language, quantum fluctuations invariably
force clocks to move occasionally "backwards in time". For
time-of-arrival operators see for example \cite{time_operators,
Kij74} and the extensive bibliography in the review \cite{MSP98}.

This discussion brings us invariably to the issue of the role of
time in quantum theory. Time enters the quantum mechanical
formalism as the {\em external} evolution parameter in
Schr\"odinger's equation; it is not an intrinsic variable of a
physical system. Indeed, outside the realm of general relativity,
time is assumed to be part of a background structure--both in
non-relativistic or special relativistic physics. For Bohr, time
is a part of the classical description of physics that is
"complementary" to the quantum description, and is needed if the
measurement theory of the later is to make sense.

In this paper we identify different candidates for the
time-of-arrival probability, by treating time as a parameters
external to the quantum system under consideration.
 This assumption  is not only forced on us by the quantum mechanical formalism,
but also corresponds to the way time is taken into account in
experiments. When an event happens (say a detector clicks) we record
the time it took place by "looking" at a clock in the laboratory.
Clocks are classical systems that do not interact with either the
quantum system or with the measurement device-- their degrees of
freedom  are  not correlated to the physical processes that take
place in the act of measurement. In effect, the time parameter we
consider is the reading of the clock that is simultaneous with the
detector's click. Since in any laboratory the relative speeds
involved are much smaller than the speed of light, there is no
problem in assuming a synchronisation of events. Hence in this
paper, we construct a probability distribution for the clock reading
that are simultaneous with the realisation of a specific quantum
event (usually a particle detection). It is important to emphasise
that the determination of the time-of-arrival involves only {\em a
single act of measurement}, which corresponds to the particle being
registered by the detector.

In each individual run of the experiment we record the moment the
particle entered the measuring device. The only way to do so is by
checking at every single moment of time which of the two
alternatives holds: the particle having been detected or not. We
then identify the time-of-arrival as the moment of transition from
the event of no-detection to the event of detection. For this
purpose, we need to consider a (continuous) history of alternatives
of detection. This suggests that a framework based on histories
(path integrals or the consistent histories approach) is
particularly suitable for that purpose. Indeed, an important feature
of the histories formalism is that it distinguishes the role of the
causal ordering of events in quantum theory from that of evolution
(see \cite{Sav99a} or the more general discussion in \cite{Sav04}).
This allows us to set up the problem of the time-of-arrival
probabilities at a kinematical level, i.e. without specific
reference to the system's Hamiltonian.

A key feature of the constructions we present here is that they are
not tied to a specific choice of the Hamiltonian that describes the
quantum system's dynamics .  From the experimentalist's point of
view, this is perhaps the most natural procedure.  One may measure
the time-of-arrival for any system without any prior knowledge about
the system's dynamics. All that is needed is a particle detector and
an external clock. We would use the same detector for a free
particle, for a particle moving in a potential, for a particle in
presence of an environment, or even when we have complete ignorance
about the particle's dynamical behaviour. The procedure one follows
to measure the time-of-arrival should not depend in principle on the
system's dynamics, only the results (namely the probabilities)
should.

 The histories description provides an
important technical advantage. While the time-of-arrival does not
define a function on the space of single-time alternatives of the
system, it becomes one in the space of histories. In fact, the
problem of defining a time-of-arrival probability is  a special
case of defining probabilities for histories. We shall exploit
this fact in section 5, in order to construct a
Positive-Operator-Valued-Measure for histories, by mirroring the
corresponding construction of probabilities in sequential
measurements.

To demonstrate unambiguously that time-of-arrival is naturally
defined in a histories framework, we consider an analogous
construction, namely  the time-of-arrival in stochastic processes
(section 2) . At this level  a probability density for the
time-of-arrival can be simply constructed. However, when we pass to
quantum theory a problem appears. The probabilities obtained from
Born's rule depend on the time variable in a way that does not allow
the straightforward definition of a probability density. As a
result, there is a strong ambiguity in the implementation of the
continuous-time limit.

The second complication involved in quantum theory is the fact
that the construction depend strongly on the interpretation of the
"wave-packet reduction" rule. One possible interpretation is that
 the wave-packet reduction is a physical process that takes place
only after the measured system has interacted with a measuring
device. Another interpretation is  that the reduction rule can be
applied to any circumstance, in which we have obtained information
about the quantum system. For example, if a particle detector did
not click at a specific moment of time, then we can infer that the
particle has been outside the region of detection at that time.

The first interpretation allows us
 to construct time-of-arrival probabilities by applying {\em classical
reasoning} to the quantum mechanical probabilities. The
corresponding probability density is not a linear functional of
the initial density matrix, because the assumption we employed
violates the "logic" of quantum mechanical propositions.
Nonetheless, it has the correct classical limit and physically
reasonable properties. Its main drawback is that it is ambiguous
with respect to the continuous-time limit, and depends on the
procedure one employs for its implementation.

The second interpretation of the reduction procedure lends itself to
constructions that emphasis the 'logic' of quantum events. All
information we obtain for a quantum system, whether this arises from
a concrete experimental datum or form inference from the lack of
such a datum (i.e. the detector not having clicked) are treated in
the same footing. An important example is the consistent histories
approach, which we examine in section 4. Time-of-arrival histories
are a special case of the so-called  "spacetime coarse-grainings"
that have been studied before by Hartle \cite{Har} and others
\cite{scc}. The advantage conferred to us by this approach is that
the continuous-time limit {\em is naturally obtained at the level of
amplitudes} (which essentially correspond to restricted path
integrals). The problem  arises at the level of combining the
amplitudes in order to obtain a probability density. In effect,
there is strong interference between different values of arrival
times that are mutually exclusive in the classical context.

The quantum Zeno effect \cite{MiSu77} poses another problem for the
determination of probabilities; it can be partially evaded but it
remains troublesome at the fundamental level. It seems that the only
information we can obtain unambiguously in this framework is the
classical deterministic limit for the time-of-arrival.

In section 5, we apply the results obtained from the analysis of
histories in a different context, namely the construction of a
Positive-Operator-Valued-Measure that  provides probabilities for
the time-of-arrival. The key idea is that the construction of a
probability density for the time of arrival is not fundamentally
different from that of probabilities for sequential measurements.
Hence, we follow a procedure developed for the study sequential
measurements \cite{Ana04, Ana05}. The resulting time-of-arrival
probabilities, like those of sequential measurements, are
contextual; they depend strongly on the resolution of the
measurement device. However, in the physically relevant regime for
the free particle the dependence on $\tau $ drops out, and the
constructed POVM coincides with that of Kijowski \cite{Kij74}.

The approach to the time-of-arrival developed in Section 5 involves
acting with the projection operator corresponding to no detection on
the system's wave-function at every moment of time. This action
loosely corresponds to the fact that we have obtained information
from the quantum system. It is important to emphasise that it does
not refer to a physical act of measurement. It cannot be described
in the language of standard quantum measurement theory: a von
Neumann measurement, for example, involves a finite time interval
during which the interaction Hamiltonian between system and
measuring device dominates over the system's self-Hamiltonian. This
is clearly not happening in the time-of-arrival set-up, at any
moment prior to the system entering the measuring device. In quantum
measurement theory, a physical measurement involves pre-measurement
and reduction, and here the former part is missing. In physical
terms, the time-of-arrival measurement involves a single measuring
device, a single act of detection, a single irreversible change in
the device and a single moment of time at which the interaction
Hamiltonian becomes dominant. The only difference from the case of
standard measurements is that the time of detection is unknown.

\section{Time-of-arrival in stochastic processes}

We first study the time-of-arrival probability in the theory of
 stochastic processes. This allows us to demonstrate the procedure
 we will follow in quantum theory, without the complications
 arising from the interpretations of the quantum measurement process.

We consider for simplicity a one-dimensional system, the state of
which is fully specified at a moment of time by the position
variable $x$. The sample space $\Omega$  is then identified with
${\bf R}$. An ensemble of such systems is described at $t=0$ by the
probability density $\rho_0(x)$. This probability density evolves
according to the law
\begin{eqnarray}
\frac{\partial \rho}{\partial t} = {\cal L} \rho, \label{stoch}
\end{eqnarray}
where ${\cal L}$ is a linear, positive, trace-preserving operator
on the space of probability densities.

We next construct the stochastic process corresponding to the
system described by Eq. (\ref{stoch}). We assume that the
measurements take place in the time interval $I = [0, T]$, where
$T$ may be eventually taken to infinity. The sample space
$\Omega^I$ for the stochastic process is the space of all
continuous paths $x(\cdot)$ from $I$ to ${\bf R}$. The relevant
random variables are the function $X_t$ on $\Omega$ defined as
$X_t(x(\cdot)) = x(t)$. There exists a probability measure on
$\Omega^I$ given by the "continuum limit" of discrete-time paths
$(t_1, x_1; t_2, x_2; \ldots; t_n, x_n)$
\begin{eqnarray}
d \mu_{\{t_1, t_2, \ldots, t_n \}}(x_1, x_2, \ldots, x_n) =
\rho_0(x_0) g(x_0,0;x_1,t_1) g(x_1, t_1;x_2, t_2) \ldots \nonumber
\\ g(x_{n-1}, t_{n-1}; x_n, t_n), dx_0 dx_1 dx_2 \ldots dx_n,
\label{prob}
\end{eqnarray}
where $g(x,t;x',t')$ represents the propagator associated to
equation (\ref{stoch}). We assume that $\rho_0$ has support
 only for values of $x < 0$.

From the probability measure (\ref{prob}) we construct the
probability for the proposition that the particle is detected at $x
= 0$ at a specific time $t$, $0< t < T$. For this purpose, we split
the interval $[0,T]$ into N time steps of width $\delta t = T/N$. We
assume also that $t = n \delta t$ for an integer $n <N$ and write $m
= N - n$. We denote by $\chi_{\pm}$ the characteristic functions of
the intervals $- \infty < x < 0$ and $0 < x < \infty$ respectively.

If the particle crosses the surface $x = 0$ for the first time
within the time-interval $[t, t+\delta t]$, then it must have been
 in the region $(-\infty, 0)$ for all times less or equal to $t$ and in the region
  $(0, \infty)$ at time $t+\delta t$. There is no
reason to make any assumption about where it will be at times larger
than $t+\delta t$, because in time-of-arrival measurements we are
interested only in the time of the first detection. The particle may
be, for example  absorbed by the detector at time $t$.

With the above considerations in mind, we see that the probability
that the particle is measured during the interval $[t, t+ \delta
t]$ equals
\begin{eqnarray}
p(x = 0| [t,t+\delta t]) = \mu(D_{[t,t+\delta t]}),
\end{eqnarray}
where $\mu$ is the stochastic probability measure and
\begin{eqnarray}
D_{[t,t+\delta t]} = \chi_- \otimes \chi_- \otimes \ldots \chi_-
\otimes \chi_+ \otimes 1 \otimes \ldots 1.
\end{eqnarray}
The function $D_{[t,t+\delta t]}$ is a characteristic function on
$\times_{t_i} \Omega_{t_1}$, and depends only on the value of $n$,
namely the time-step that corresponds to detection. We may then
also write $D_{[t,t+\delta t]}$ as $D_n$. If we also define by
$\bar{D}$ the characteristic function
\begin{eqnarray}
\bar{D} = \chi_- \otimes \chi_- \otimes \ldots \otimes \chi_-
\end{eqnarray}
that corresponds to the particle never crossing $x = 0$ within the
time interval $[0,T]$, the following relations hold
\begin{eqnarray}
D_n D_m &=& D_n \delta_{nm} \\
D_n \bar{D} &=& 0 \\
\sum_{n=0}^N D_n + \bar{D} &=& 1.
\end{eqnarray}
The variables $D_n, \bar{D}$ then define an exclusive and exhaustive
set of alternatives\footnote{This decomposition is a special case of
the  Path-Decomposition-Expansion and has been used in the context
of restricted path-integrals in reference \cite{Hal95}.}, hence the
restriction of the probability measure to the algebra they generate
defines a proper normalised probability measure for the
time-of-arrival in discrete time.

We next construct the continuous limit of this probability as $N
\rightarrow \infty$. We use an operator notation, representing the
action of the integral kernel as $e^{{\cal L}t}$. Using Eq.
(\ref{prob}) we write
\begin{eqnarray}
p(x = 0| [t,t+\delta t]) = \int dx [\chi_+ e^{{\cal L} \delta
t}[\chi_- e^{{\cal L}\delta t}]^n\rho_0](x)
\label{discrete-stoch}.
\end{eqnarray}
If we denote by $K_t$ the limit
\begin{eqnarray}
K_t = \lim_{n \rightarrow \infty} [\chi_- e^{{\cal L} t/n}
\chi_-]^n,
\end{eqnarray}
we obtain
\begin{eqnarray}
p(x = 0| [t,t+\delta t]) = \delta t \int dx [\chi_+ {\cal L}
\chi_-(x) K_t \rho_0](x)
\end{eqnarray}
The fact that the probability of the first crossing is
proportional to $\delta t$ implies that we can pass to the
continuum limit defining a probability density on $[0,T]$
\begin{eqnarray}
p(t|x=0) = \int dx [\chi_+ {\cal L} K_t \rho_0](x).
\label{classdensity}
\end{eqnarray}
This probability density is not normalised to one as there is a
no-zero residual probability $p(N)$ that the particle is not
detected at all within $[0,T]$
\begin{eqnarray}
\int_0^T dt \, p(t|x=0) = 1 - \int dx [ K_T \rho_0](x) : = 1 - p(N)
\end{eqnarray}
For generic initial states and dynamics the residual probability
does not vanish as $t \rightarrow \infty$.

Using the probability density $p(t|x=0)$, we  define the probability
of detection within any interval $[t_1, t_2]$ by integrating
$p(t|x=0)$ in this interval. The definition of average values of
quantities is slightly more intricate. The sample space for the time
of arrival (at the continuum limit) is not the interval $[0,T]$, but
the set $[0,T] \cup \{N \}$, where $N$ refers to the event of no
detection. Strictly speaking,
 physical observables are functions on $[0,T] \cup \{ N
\}$. Hence there is an ambiguity in the definition of a function
representing time $t$, because there is no natural numerical value
it can take when evaluated on $N$.

In effect, a time function is defined unambiguously as a {\em
conditional expectation}, namely after the assumption that the
particle has actually been detected. This implies that we restrict
(condition) the sample space to  $[0, T]$. The conditional
probability $p_c$ density is then
\begin{eqnarray}
p_c(t|x=0) = \frac{p(t|x=0) }{1 - p(N)}.
\end{eqnarray}

\subsection{Examples}

\subsubsection{Two level system.} In the derivation of the
probability of time-of-arrival, we referred to the variable $x$ as
position. However, the derivation is completely general and Eq.
(\ref{density}) may be applied to any sample space. If the latter is
discrete, we have to exchange the integral with a summation. We may
consider for example a stochastic two-level system (a bit) and
determine the probability for the time-of-transition. In this case
the sample space then consists of two alternatives: $0$ and $1$. We
assume that initially the system is found at state $0$ . The most
general operator ${\cal L}$ consistent with positivity and
normalisation of probabilities is
\begin{eqnarray}
{\cal L} = \left( \begin{array}{c c} -a&  b \\ a & -b \end{array}
\right)
\end{eqnarray}
The corresponding transition matrix for a small time interval
$\delta t$ is
\begin{eqnarray}
 \left( \begin{array}{c c} 1 -a \delta t&  b \delta t \\
a \delta t& 1 -b \delta t \end{array} \right),
\end{eqnarray}
 which is the most
general stochastic map for a two-level system.

It is then easy to compute  the probability density for the time
of the transition $0 \rightarrow 1$
\begin{eqnarray}
p(0\rightarrow 1;t) = b e^{-bt}.
\end{eqnarray}
The probability  $p(N)$ that no transition took place within the
time interval $[0,T]$ equals $e^{- b T}$. As $T \rightarrow
\infty$, $p(\infty) = 0$ and we may compute the mean time of
transition
\begin{eqnarray}
<t> = b^{-1},
\end{eqnarray}
which are the standard results for decay processes.

\subsubsection{Wiener process}

 We next consider the case of the
Wiener process, defined by the evolution operator
\begin{equation}
{\cal L} \rho = \frac{D}{2} \partial^2 \rho,
\end{equation}
where $D$ is a diffusion constant. We assume that the particle is
initially localised at $x = - L$, namely $\rho_0(x) =
\delta(x+L)$.

The operator $K_t$ is the propagator corresponding to ${\cal L}$
with the Dirichlet boundary conditions at $x = 0$ \footnote{ This
is in fact a more general result. A quick but not fully rigorous
way to see this is by writing the
 characteristic function of a set $C$ as
$\chi_C(x) = e^{-V_C(x) \delta t}$, where $V_C(x)$ is a "confining
potential" that takes value $0$ within $C$ and $\infty$ outside
$C$. We may then use the Trotter product formula $\lim_{n
\rightarrow \infty} (e^{{\cal L}t/n} e^{-V_- t/n})^n = ^{({\cal L}
- V_-)t}$, which is exactly the propagator for ${\cal L}$ with
Dirichlet  conditions at the boundaries of $C$.}.

The integral kernel $K(x,x';t)$ corresponding to $K_t$ is
\begin{eqnarray}
K(x,x';t) = \chi_-(x) \chi_-(x') \sqrt{\frac{1}{2\pi Dt}}
\left(e^{-(x-x')^2/2Dt} - e^{-(x+x')^2/2Dt} \right),
\end{eqnarray}
yielding
\begin{eqnarray}
p(t|x=0) = \sqrt{\frac{1}{2\pi Dt}} \frac{L}{2t} e^{-L^2/2Dt},
\end{eqnarray}
while
\begin{eqnarray}
p(N) =   erf(L/\sqrt{2DT}) - erf(-L/\sqrt{2DT}),
\end{eqnarray}
where $erf$ is the error function.

\section{The Copenhagen description}

\subsection{The standard construction and its problems}

We next  attempt to construct the time-of-arrival probability for
quantum theory  within the Copenhagen interpretation.

 We split the time-interval $[0,T]$ into
$n$ time-steps of width $\delta t = T/n$  We represent the
projection operators that correspond to the particle lying within
$(-\infty, 0]$ and in $[0, \infty)$  as $\hat{P}_-$ and $\hat{P}_+$
respectively, The time-of-arrival is defined as the moment the
particle crosses from $(- \infty, 0)$ to $[0, \infty)$. We assume
that at $t = 0$ the particle is described by a density matrix
$\hat{\rho}_0$.

The probability that the particle crossed $x=0$ at time $t_1$
equals $p_1 = Tr(\hat{\rho} e^{i\hat{H}t_1} \hat{P}_+ e^{-i
\hat{H} t_1})$. The probability that the particle crossed $x=0$ at
the next moment $t_2$ is then equal to
\begin{eqnarray}
p_1 \; p(-, t_1;+, t_2),
\end{eqnarray}
where $p(-,t_1;+ t_2)$ is the conditional probability that the
particle was found within $(-\infty, 0]$ at $t_1$ and within $(0,
\infty)$ at $t_2$. According to the standard "reduction" rule this
equals
\begin{eqnarray}
p(-,t_1;+ t_2) = \frac{ Tr (\hat{P}_+ e^{-i \hat{H}(t_2 - t_1)}
\hat{P}_- e^{-i\hat{H}t_1} \hat{\rho}_0 e^{i \hat{H}t_1} \hat{P}_-
e^{i \hat{H}(t_2 - t_1)} )}{Tr(\hat{\rho} e^{i\hat{H}t_1}
\hat{P}_- e^{-i \hat{H} t_1})}, \label{conditional}
\end{eqnarray}
yielding
\begin{eqnarray}
p_2 = Tr (\hat{P}_+ e^{-i \hat{H}(t_2 - t_1)} \hat{P}_-
e^{-i\hat{H}t_1} \hat{\rho}_0 e^{i \hat{H}t_1} \hat{P}_- e^{i
\hat{H}(t_2 - t_1)} ).
\end{eqnarray}
Following the same procedure we obtain the probability that the
particle crossed $x = 0$ at the $k$-th time step
\begin{eqnarray}
p_k = Tr( \hat{P}_+e^{-i \hat{H}(t_k -t_{k-1})}\ldots\hat{P}_+
e^{-i \hat{H}(t_2 - t_1)} \hat{P}_- e^{-i\hat{H}t_1} \hat{\rho}_0
\nonumber \\ \times e^{i \hat{H}t_1} \hat{P}_- e^{i \hat{H}(t_2 -
t_1)} \hat{P}_- \ldots e^{i \hat{H}(t_k -t_{k-1})}).
\end{eqnarray}
As we take the continuum limit $n \rightarrow \infty$, assuming
that the initial state has support only in $( - \infty, 0)$, we
obtain the probability that the particle was found between $t$ and
$t+ \delta t$
\begin{eqnarray}
p([t,t+\delta t]) = \delta t^2 Tr ( (\hat{C}_t \hat{H}
\hat{\rho}_0 \hat{C}^{\dagger}_t \hat{H} \hat{P}_+  ),
\end{eqnarray}
where
\begin{eqnarray}
\hat{C}_t = (\hat{P}_- e^{-i \hat{H} t/n} \hat{P}_-))^n.
\label{Ct}
\end{eqnarray}

There are two severe problems in this result. First, the probability
$p([t,t+\delta t])$ is proportional to $\delta t^2$, and hence does
not define a probability density. If we tried to construct the
probability for the detection within a finite time interval $[t_1,
t_2]$ by integration, we would, strictly speaking, obtain zero, i.e.
we would  find that the particle is never detected.

Second, the probability that the particle never crosses $x = 0$
within the time interval $[0, T]$ equals
\begin{eqnarray}
p(N) = Tr (\hat{C}_T \hat{\rho}_0 \hat{C}^{\dagger}_T).
\end{eqnarray}
The operator $\hat{C}_t$ is a degenerate unitary operator with a
support in the range of $\hat{P}_-$ \cite{MiSu77}--this is  the
well-known quantum Zeno paradox. It follows that $\hat{C}_t
\hat{C}^{\dagger}_t = \hat{P}_-$, hence
\begin{eqnarray}
p(N) = Tr (\hat{\rho}_0 \hat{P}_-) = 1.
\end{eqnarray}

It seems as though the particle can never cross $x = 0$, which is
clearly a mistake. This problem can be addressed by noticing that
the probabilities $p([t,t+\delta t])$ are by definition
non-additive with respect to the projectors and hence do not
satisfy the Kolmogorov additivity condition, i.e. they are not
probabilities at all. It is therefore no surprise that they  do
not define a probability density. One should employ for the
continuum limit a probabilistic object that is additive with
respect to the projectors. This is the decoherence functional of
the consistent histories approach. The problem of the quantum Zeno
effect is also partially alleviated in consistent histories, since
the quantity $p(N)$ is not a genuine probability, unless a
decoherence condition is satisfied. There are  other problems,
however, that appear in the implementation of  the theory's
classical limit. We will discuss this issue in more detail in the
next section.

\subsection{A different interpretation of the reduction procedure}

An objection that can be raised to the derivation above has to do
with the use of the conditional probability rule in Eq.
(\ref{conditional}). One may argue that since the particle has not
been detected at time $t_2$, it has not interacted with the
measuring device and for this reason,  one should not employ the
"reduction of the wave packet" rule, because no measurement has
actually taken place\footnote{See the discussion in the
Introduction.}. Instead of the conditional probability, one should
use the full probability of detection at $t_2$, namely
\begin{eqnarray}
 Tr (\hat{P}_+ e^{-i\hat{H} t_2} \hat{\rho}_0 e^{i
\hat{H}t_2}) := Tr (\hat{P}_+ \rho_{t_2})
\end{eqnarray}
hence the probability of detection at the $k$-th time step equals
\begin{eqnarray}
p_k = (1 - \sum_{i=0}^{k-1} p_i) Tr(\hat{\rho}_{t_k} \hat{P_+}),
\end{eqnarray}
This is a recursive equation with the following solution
\begin{eqnarray}
p_k = Tr(\hat{\rho}_{t_k} \hat{P}_+) \prod_{-=0}^{k-1} [1 -
Tr(\hat{\rho}_{t_i} \hat{P_+})] \nonumber
\\
= Tr(\hat{\rho}_{t_k} \hat{P}_+)\exp \left[ \sum_{i=0}^{k-1} \log
( 1 -Tr(\hat{\rho}_{t_i} \hat{P_+})) \right]. \label{operdiscrete}
\end{eqnarray}

The expression above does not have a natural continuum limit. The
sum in the exponential does not define an integral, because a
$\delta t$ term is missing. Again, we face the problem that the
dependence of the quantum mechanical probabilities on time does not
correspond to the existence of a genuine probability measure--hence
a continuous limit does not exist naturally.

One way to obtain a probability measure is to introduce a time-step
$\tau$, which is a measure of the temporal resolution of the
measuring device. For any time-scales much larger than $\tau$ one
may substitute the sum in equation ({\ref{operdiscrete}) with an
integral, thus obtaining a probability measure on $[0, T]$
\begin{eqnarray}
p(t) = \frac{1}{\tau} Tr (\hat{\rho_t} \hat{P}_+) \exp[
\frac{1}{\tau} \int_0^t ds Tr( \hat{\rho}_s \hat{P}_-) ] =
\nonumber \\
= - e^{t/\tau} \frac{d }{d t}e^{-F(t)}, \label{opercont}
\end{eqnarray}
where $F(t) = \frac{1}{\tau} \int_0^t ds \hat{\rho}_s \hat{P}_-$.

The probability density above has a physically reasonable behaviour
for a time-of-arrival probability. For a wave function, whose center
follows approximately a classical path, the probability density
(\ref{opercont}) is peaked around the classical time-of-arrival for
this path. To see this one may consider Eq. (\ref{opercont}) for a
free particle of mass $m$. Considering an initial Gaussian state
\begin{eqnarray}
\psi_0(x) = \frac{1}{[\pi \sigma^2]^{1/4}} e^{ - \frac{(x +L)^2}{2
\sigma^2} + i p x}
\end{eqnarray}
peaked at $t=0$ around $x = - L$ with mean momentum $p$, we obtain
for the function $F(t)$
\begin{eqnarray}
F(t) = \frac{1}{2\tau} \int_0^t ds \left[ 1 + Erf \left(\frac{(L -
\frac{p}{m}t)}{\sqrt{\sigma^2 + \frac{t^2}{m^2 \sigma^2}}}\right)
\right],
\end{eqnarray}
which implies that (\ref{opercont}) has a strong peak around the
classical time-of-arrival $t_{cl} = \frac{Lm}{p}$.

 The problem lies in the strong dependence of these
probabilities on the parameter $\tau$. While it is reasonable to
assume that the probabilities will be dependent on parameters that
characterise the method of detection, we would intuitively expect
that this dependence would be insignificant when we consider
sufficiently large intervals of time. This is definitely not the
case here as the probabilities are very sensitive on the value of
$\tau$.

Since quantum theory does not provide a natural way to pass into the
continuum limit (at least in the scheme we consider in this
section), it is natural to expect that different procedures will
lead to different results. We may consider for example the following
alternative.

 In Eq. (\ref{operdiscrete}) we may substitute in place of
  $\hat{P}_+$ a projector $\hat{P}_{\delta x}$   in position of width $ \delta x$
around $x=0$, and the projector $\hat{1} - \hat{P}_{\delta x}$ in
place of $\hat{P}_-$ . This corresponds to a set-up by which the
particle is detected only if it crosses the region $[-\delta x/2,
\delta x/2]$. We next assume that the size $\delta x$ decreases with
$\delta t$, so that as $\delta t \rightarrow 0$, the area of
detection also shrinks to zero. We therefore  write $\delta x = v
\delta t$, for some unspecified constant $v$ with dimensions of
velocity. This way of taking the limit essentially implies that
actual detection of a particle needs a finite time-interval, since
at the limit $\delta t \rightarrow 0$, $\hat{P}_{\delta x} =
\hat{1}$.

Writing $\rho_t(0) = \langle x = 0 | \hat{\rho}_t |x = 0 \rangle$,
we obtain at the limit $\delta t \rightarrow 0$ the probability
that the particle is detected between time $t$ and $t+\delta t$ as
\begin{eqnarray}
p(t) \, \delta t = \delta t \, v \, \rho_t(0) e^{- v \int_0^t ds
\rho_s(0)} = - \delta t \frac{d}{d t} e^{ - v \int_0^t ds
\rho_s(0)}. \label{density}
\end{eqnarray}
Hence, the probability that the particle is detected within the
time interval $[t_1, t_2]$ equals
\begin{eqnarray}
p([t_1, t_2]) = e^{ -  v \int_0^{t_1} ds \rho_s(0)} - e^{ - v
\int_0^{t_2} ds \rho_s(0)},
\end{eqnarray}
while the probability that the particle is not detected within the
time-interval $[0, T]$ equals
\begin{eqnarray}
p(N)  = e^{ -  v \int_0^{T} ds \rho_s(0)}.
\end{eqnarray}

The probability density (\ref{density}) has the correct behaviour at
the classical limit, but again it depends on an unspecified
parameter $v$, which this time has dimensions of velocity. One has
to assume that $v$ has to be identified with a characteristic
property of the measuring device.

It follows that with the interpretation of the reduction rule we
employed here, a probability distribution for the time-of-arrival
cannot be constructed without making reference to the specific
set-up through which it is determined. Whatever scheme one might
employ, one has to introduce additional parameters in the
description.

It would be mistaken, however, to consider the derivation leading to
Eqs. (\ref{density}) or (\ref{opercont}) as inherently faulty. The
only assumption we employed is that the reduction rule can only be
applied, when an actual measurement has actually taken place, and
not when we make an {\em inference} about the system by the fact
that no detection has occurred. This implies, in particular, that
the quantum Zeno effect is irrelevant for the time-of-arrival,
because we do not have a continuous act of measurement (only a
continuous inference). With this assumption the proof leading to
(\ref{operdiscrete}) only employs the classical rules of probability
theory. In that sense, the key mathematical problem is that the
dependence of the quantum mechanical probabilities on time does not
allow the definition of a stochastic process--see the related
discussion in \cite{Ana05}. Quantum probabilities {\em are not
naturally densities with respect to time}; one can make them
densities by introducing additional parameters.

It is important to note that this fundamental difficulty does not
go away, when we enlarge the system by including degrees of
freedom of the measurement device. The problem of finding a
suitable continuous-expression for (\ref{operdiscrete}) does not
depend on specific features of the system's Hilbert space. The
density matrix may include degrees of freedom of the measuring
device or of an environment. The problem lies with the way time
appears in the formalism of quantum theory.

In any case, equations (\ref{density}) and (\ref{opercont})
provide interesting candidates for a probability distribution for
a time-of-arrival. They have a proper classical limit and  they
are mathematically unambiguous. In principle, they could be
checked by any precision measurement of times-of-arrival.

\section{The histories description}

In this section, we follow a different approach from that of section
3.2. We assume that the reduction rule can  be applied in any case
we have obtained information about a quantum system. This allows us
to preserve the `logical' structure of quantum mechanical
propositions. The natural scheme to explore the time-of-arrival
problem is then the consistent histories approach. However, the
results we obtain here are of a more general character. The
mathematical objects employed in the consistent histories approach
are essentially path-integrals and the amplitudes defined by these
path integrals can be employed for the study of the time-of-arrival
in different schemes. (We do that in section 5). The most important
gain from this approach is that the continuous-time limit can be
obtained unambiguously, because it is implemented at the level of
amplitudes and not at that of probabilities.

\subsection{Consistent histories }

   The consistent histories approach to quantum theory was  developed by
Griffiths \cite{Gri84}, Omn\'es \cite{Omn8894}, Gell-Mann and
Hartle \cite{GeHa9093, Har93a}. The basic object is a
    history, which   corresponds to  properties of the physical
system at successive instants of time. A discrete-time history
$\alpha$ will then correspond to a string $\hat{P}_{t_1},
\hat{P}_{t_2}, \ldots \hat{P}_{t_n}$ of projectors, each labelled
by  an instant  of time. From them, one can construct the class
operator
\begin{equation}
\hat{C}_{\alpha} = \hat{U}^{\dagger}(t_1) \hat{P}_{t_1}
\hat{U}(t_1) \ldots \hat{U}^{\dagger} (t_n) \hat{P}_{t_n}
\hat{U}(t_n)
\end{equation}
where $\hat{U}(s) = e^{-i\hat{H}s}$ is the time-evolution
operator. The probability for the realisation of this history is
\begin{equation}
p(\alpha) = Tr \left( \hat{C}_{\alpha}^{\dagger}\hat{\rho}_0
\hat{C}_{\alpha} \right), \label{decfundef}
\end{equation}
 where $\hat{\rho}_0$ is the density matrix describing the system at time $t = 0 $.

However, the expression above does not define a probability measure
in the space of all histories, because the Kolmogorov additivity
condition cannot
 be satisfied: if $\alpha$ and $\beta$ are exclusive histories, and $\alpha \vee \beta$
denotes their conjunction as propositions, then it is not true
that
\begin{equation}
p(\alpha \vee \beta ) = p(\alpha) + p(\beta) .
\end{equation}
 The
histories formulation of quantum mechanics does not, therefore,
enjoy the status of a genuine
 probability theory.

However,  an
 additive probability measure {\it is} definable, when we restrict to
particular  sets of histories.
 These are called {\it consistent sets}. They are more conveniently
defined through the introduction of a new object: the decoherence
functional. This is a complex-valued function of a pair of
histories given by
\begin{equation}
d(\alpha, \beta) = Tr \left( \hat{C}_{\beta}^{\dagger}
\hat{\rho}_0 \hat{C}_{\alpha} \right).
\end{equation}
A set of exclusive and exhaustive alternatives is called
consistent, if for all  pairs of different histories $\alpha$ and
$\beta$, we have
\begin{equation}
 Re \hspace{0.2cm} d(\alpha, \beta) = 0 .
\end{equation}
In that case one can use equation (2.2) to assign a probability
measure to this set.

\subsection{Time-of-arrival histories}

Histories and propositions about histories may be represented by
projection operators on a Hilbert space ${\cal V} = \otimes_t
H_t$, which the tensor product of the single time Hilbert spaces
of standard theory--this is the History Projection Operator (HPO)
formulation of the history theory \cite{I94}. The merit of this
description is that the logical structure of history propositions
is preserved (they forma lattice that corresponds with the lattice
of subspaces of ${\cal V}$), and in the present context allows the
arguments used for the time-of-arrival description of stochastic
processes to be transferred into the quantum level. In particular,
the continuum limit in time may be taken in an unambiguous manner.
Note that in this scheme the decoherence functional is a
Hermitian, bilinear functional on ${\cal V} \times {\cal V}$ that
satisfies the following properties
\begin{eqnarray}
d(1,1) =1 \\
d(0, \alpha) = 0 \\
d(\alpha, \alpha) \geq 0
\end{eqnarray}

 We next consider  a
description of time-of-arrival histories with $n$-time steps. One
defines the projectors $\hat{\alpha}_n$ corresponding to the
proposition that the particle crossed $x=0$ for the first time
between the $m$-th and the $m+1$-th time step
\begin{eqnarray}
\hat{\alpha}_m = \hat{P}_-   \otimes \hat{P}_- \otimes \ldots
\otimes \hat{P}_+ \otimes \hat{1} \otimes \ldots \otimes \hat{1},
\label{history}
\end{eqnarray}
as well as the projector $\hat{\bar{\alpha}}$ corresponding to the
proposition that the particle does not cross $x = 0$ within the
$n$- time steps
\begin{eqnarray}
\hat{\bar{\alpha}} = \hat{P}_- \otimes \hat{P}_- \otimes \ldots
\otimes \hat{P}_- \label{history2}
 \end{eqnarray}
 Clearly these projectors satisfy
 \begin{eqnarray}
\hat{\alpha}_n \hat{\alpha}_m &=& \delta_{nm} \hat{\alpha}_n \\
\hat{\alpha}_n \hat{\bar{\alpha}} &=& 0 \\
\sum_m \hat{\alpha}_m + \hat{\bar{\alpha}} &=& 1.
 \end{eqnarray}
Thus they form a exhaustive and exclusive set of histories, hence a
sublattice of the lattice of history propositions\footnote{One
should note that the $n$-time histories we study here should not be
viewed as discretisations of continuous time paths, but as histories
corresponding to genuinely discrete time. The consideration of
discretized alternatives in a continuous-time theory is conceptually
problematic because at any time between $t_i$ and $t_{i+1}$ the
particle may have crossed $x = 0$, and this fact will not be
captured in the resulting propositions. Our approach is that we
first consider alternatives of detection in a discrete-time theory,
and we then identify a suitable continuous limit for the decoherence
functional. This involves a choice on the way we {\em define} the
continuous histories. This choice allows us to recover known results
\cite{Har, Facchi}. However, this procedure may not be unique--see
the discussion on alternative treatments of the quantum Zeno effect.
}. One can therefore pullback the decoherence functional to this
lattice, thereby obtaining a decoherence functional on a sample
space consisting of the points $(t_1, \ldots, t_n)$ together with
the point $N$ corresponding to no crossing
\begin{eqnarray}
d(t_n, t_m) &=& d(\hat{\alpha}_n, \hat{\alpha}_m) \\
d(N, t_n ) &=& d(\hat{\bar{\alpha}}, \hat{\alpha}_n) \\
d(N,N) &=& d(\hat{\bar{\alpha}}, \hat{\bar{\alpha}})
\end{eqnarray}

In analogy to the stochastic case, one may define a self-adjoint
time-of-arrival operator $\hat{T}$ on ${\cal V}$ modulo its value on
the subspace corresponding to $\hat{\bar{\alpha}}$, namely one may
define
\begin{eqnarray}
\hat{T}_{x=0} = \sum_i t_i \hat{\alpha}_i,
\end{eqnarray}
which is unambiguously defined on ${\cal V} -
Ran(\hat{\bar{\alpha}})$ -- $Ran(\hat{\bar{\alpha}})$ is the
closed linear subspace corresponding to $\hat{\bar{\alpha}}$.

We next consider
  two discretisations $\{t_0 =0, t_1, t_2, \ldots t_N = T\}$ and $ \{t'_0
=0, t'_1, t'_2, \ldots t'_{N'} = T \}$ of the time interval $[0.T]$
with time-step $\delta t = T/N$, and $\delta t' = T/N'$. We
construct the decoherence functional $d([t, t+\delta t], [t',
t'+\delta t'])$,  where $ n = t N/T$ and $m = t' N'/T$. This reads
\begin{eqnarray}
d([t, t+\delta t], [t', t'+\delta t']) = Tr \left( \hat{\rho}_0
[e^{i \hat{H} \delta t'} \hat{P}_-]^n e^{i \hat{H}\delta t'}
\hat{P}_+ \right. \nonumber \\
\left. \times e^{i \hat{H}(t'-t)} \hat{P}_+ e^{-i \hat{H} \delta
t} [\hat{P}_- e^{-i \hat{H} \delta t}]^m \right).
\end{eqnarray}
We take then the limit $N, N' \rightarrow \infty$, while keeping
$t$ and $t'$ fixed. Assuming that $\rho_0$ lies within the range
of $\hat{P}_-$, i.e. $\hat{P}_- \hat{\rho}_0 \hat{P}_- =
\hat{\rho}_0$ we obtain
\begin{eqnarray}
d([t, t+\delta t], [t', t'+\delta t']) = \delta t \delta t' Tr
\left(   e^{-i \hat{H}(t'-t)} \hat{P}_+ \hat{H} \hat{P}_-
\hat{C}^{\dagger}_{t} \hat{\rho}_0 \hat{C}_{t'} \hat{P}_- \hat{H}
\hat{P}_+  \right),
\end{eqnarray}
where $\hat{C}_t = \lim_{n \rightarrow \infty} (\hat{P}_- e^{-i
\hat{H} t/n} \hat{P_-})^n$.  Writing
\begin{eqnarray}
\rho(t,t') = Tr \left(   e^{-i \hat{H}(t'-t)} \hat{P}_+ \hat{H}
\hat{P}_- \hat{C}^{\dagger}_{t'} \hat{\rho}_0 \hat{C}_{t}
\hat{P}_- \hat{H} \hat{P}_+  \right) \label{densitydecf}
\end{eqnarray}
we see that the decoherence functional corresponds to a
complex-valued density on $[0,T] \times [0,T]$. The additivity of
the decoherence functional (which reflects the additivity of
quantum mechanical amplitudes)  allows us to obtain a continuum
limit, something that could not be done if we worked at the level
of probabilities.
 Consequently, one may obtain
its values on coarse-grained histories corresponding to
time-of-arrival lying within the subsets $[t_1,t'_2]$ and $[t_1',
t'_2]$ of $[0, T]$ by integrating over $\rho(t,t')$
\begin{eqnarray}
d([t'_1, t'_2], [t_1, t_2]) = \int_{t_1}^{t_2} dt \int
_{t'_1}^{t'_2} dt' \rho(t,t')
\end{eqnarray}

We then obtain the values of the decoherence functional for any pair
of measurable subsets of $[0, T]$. However, the decoherence
functional on $[0, T]$ is not properly normalised, because the
actual space of time-of-arrival propositions is not the space of
subsets of $[0, T]$, but rather the space of subsets of $[0,T] \cup
\{N\}$, where $N$ corresponds to the event of no detection. The
values of the decoherence functional, when at least one of its
entries is $N$ are easily computed
\begin{eqnarray}
d([t_1, t_2], N) &=& \int_{t_1}^{t_2} dt \; \; Tr  \left(   e^{-i
\hat{H}(T-t)}  \hat{C}^{\dagger}_{T} \hat{\rho}_0 \hat{C}_{t}
\hat{P}_- \hat{H} \hat{P}_+  \right) \\
d(N,N) &=& Tr(\hat{C}^{\dagger}_T \hat{\rho}_0 \hat{C}_T ) = 1
\end{eqnarray}
The last equation is due to the fact that the operator $\hat{C}_t$
is a degenerate unitary operator with support on the range of
$\hat{P}_-$ (the quantum Zeno effect).

The normalisation condition for the decoherence functional implies
that
\begin{eqnarray}
d([0,T], [0,T]) +  d([0, T], N) + d(N, [0, T]) + d(N, N) = 1,
\end{eqnarray}
which leads to
\begin{eqnarray}
d([0,T], [0,T])  = - 2 Re \, d([0, T], N)
\end{eqnarray}

It is important to note that in the context of consistent histories
the fact that $d(N, N) = 1$ does not imply that the event $N$ (never
crossing $x = 0$) will be realised, because $d(N,N)$ does not
correspond to a probability, unless the consistency condition $Re
\,d[0, T], N) = 0$ is satisfied. In this case, however,   $d([0, T],
[0, T]) =0$ and hence that crossing $x = 0$ never takes place.  This
implies that the event of the particle not crossing the surface $x =
0$ can only be a member of a consistent set, in which the
probability for crossing $x = 0$ is zero. This is rather
counterintuitive,  because it fails to give a correct classical
limit--see related discussion in \cite{Yam96}. One would expect that
at some level of coarse-graining one would obtain the classical
result, namely a probability distribution sharply peaked around the
classical time of arrival $t_{cl}$, similar to the one we derived in
the last section.

\paragraph{Possible treatments of the quantum Zeno effect.}

A key feature of  the quantum Zeno effect is that it is not robust.
When one employs  a positive operator $\hat{E}$ in the definition of
the operator $\hat{C}_t$ (instead of a projector),the result is no
more a degenerate unitary operator. This is true even if the
operator $\hat{E}$ is very close to a true projector $Tr|\hat{E}^2 -
\hat{E}| = \epsilon /Tr \hat{E} $, for a number $\epsilon << 1$. In
that case the matrix elements of $\hat{C}_t$ fall with $e^{-
\epsilon t}$, as we demonstrate in a simple example in the Appendix.
This implies that even a very small deviation from a true projector
leads to a qualitatively different result.

If our calculation of probabilities refers to actual measurements
then the quantum Zeno effect should not pose a problem. Realistic
measurements should be represented by POVMs rather than
projection-valued-measures, in which case the quantum Zeno effect
does not arise. However, this would spoil the continuous-time limit,
which depends crucially on the fact that the operators $\hat{P}_+$
and $\hat{P}_-$ correspond to exclusive alternatives. A naive
substitution of the operators $\hat{P}_+$ and $\hat{P}_-$, by
partially overlapping approximate projectors would introduce extra
terms in the expression of the decoherence functional. It is easy to
verify that these terms would be of the order of $||\hat{P}_+
\hat{P}_-||$ and not dependent on $\delta t$, hence they would
remain non-zero even at  the limit $\delta t \rightarrow 0$. One
could entertain the possibility that the approximate projectors
could be dependent on $\delta t$, and that at the limit of $\delta
t$ their overlap becomes zero, i.e. they become true projectors. We
have explored this possibility, but it does not seem to work. The
Zeno effect still persists at the continuum limit. (This can easily
be seen in the example we provide in the Appendix.)

While the continuous-time limit we constructed here leads
invariably to a quantum Zeno effect, this is not the only way that
this limit can be taken in the histories formalism.
 The construction of
the decoherence functional we presented here is obtained from a
limit of discrete time expression. We have not constructed actual
continuous time histories and defined the decoherence functional
straightforwardly on them. To do that one should proceed in the
logic of the HPO approach and construct a history Hilbert space that
would correspond in some sense to a "continuous-tensor product" of
single-time Hilbert spaces. Such Hilbert spaces have been
constructed before \cite{continuous_time, Sav99a}; they are not
genuine continuous tensor products, but they share many of their
features, and they are obtained from group-theoretical arguments. A
key property of this construction is that propositions have support
on finite time-intervals $[t_1, t_2]$ and not on sharp points of the
real line. The operator structure is then quite different at the
kinematical level, and it raises the possibility that one could
define a decoherence functional as a genuinely continuum object, in
a way that does not suffer from the Zeno effect. For example, it is
plausible that the operators entering the definition of the operator
$\hat{C}_t$ as the limit of $\delta t \rightarrow 0$, should also be
dependent on $\delta t$, as they should refer to finite intervals of
the real line rather than sharp points. As we argued earlier even a
small change might be sufficient to remove the undesirable
properties of $\hat{C}_t$; the real issue is to justify such changes
from first principles within the continuous-time-histories
formalism. We shall elaborate on this construction in a follow-up
paper.

\paragraph{Conditioning.} As we showed in section 2, it is
possible in classical probability
 to reduce the
probability measure from the full algebra of subsets in $[0, T]
\times \{N\}$ to the algebra of events on $[0, T]$. This reduction
results from the use of conditional probability. We defined a
probability for the time-of-arrival conditioned upon the premise
that the particle did cross $x = 0$ at some time within the
interval $[0, T]$.

This reasoning may be partially transferred to the quantum case.
We cannot speak, however, for a conditional probability because
this involves the consideration of consistent sets. Classical
conditional probability is defined through a natural mathematical
procedure, which employs the additivity of the probability measure
over the space of functions on the sample space, to reduce the
level of description into a subalgebra of events. Quantum
probabilities are not additive over the lattice of events, but the
decoherence functional is. In reference \cite{Ana03} the procedure
of conditioning at the level of decoherence functional has been
developed in detail, by generalising the classical notion of
conditional expectation to the quantum level. One may define a
decoherence functional over a subalgebra of events (namely
propositions about histories), thus incorporating any information
we may have obtained for the system. For the details of the
procedure we refer the reader to \cite{Ana03}, but for the simple
case that
 the subalgebra with respect to which we implement the conditioning is generated by
a single history proposition $\beta$, such that $d(\beta, \beta)
\neq 0$, the conditioned decoherence functional is given by the
intuitively simple expression

\begin{eqnarray}
d_c(\alpha, \alpha') =  \frac{d(\alpha \wedge\beta, \alpha' \wedge
\beta_i)}{ d(\beta, \beta)}
\end{eqnarray}

 The resulting decoherence functional is
the mathematical object that should be used in the derivation of
probabilities, provided we know that events corresponding to the
subalgebra have been realised.

In the present case, we need to condition the decoherence
functional from the algebra of events corresponding to the sample
space $[0, T] \times \{N\}$ to the subalgebra of events
corresponding to a sample space $[0, T]$, namely assuming that the
particle has actually been detected . Since this operation
involves "discarding" only a simple point of the sample space, the
result is very simple. The conditioned decoherence functional
$d_c$ is obtained by a conditioned density
\begin{eqnarray}
\rho_c(t,t') = \frac{\rho(t,t')}{d([0,T], [0,T])} =
\frac{\rho(t,t')}{- 2 Re \;d([0,T], N)}.
\end{eqnarray}
Since the event of no detection is removed from the resulting
subalgebra, we may pretend that we have avoided the quantum Zeno
effect. This is, however,  an evasion and not a solution to the
problem. It only allows one to differentiate the problem of
defining a probability for the time-of-arrival, from the more
general issue of properly defining a continuum limit, that avoids
the quantum Zeno effect.

\subsection{The free particle}

For the simple case of a particle at a line with Hamiltonian
$\hat{H} = \frac{\hat{p}^2}{2 M} + V(\hat{x})$, where the potential
is bounded from below, we may employ a result in \cite{Har, Facchi}
that the restricted propagator $\hat{C}_t$ is obtained from the
Hamiltonian $\hat{H}$ by Dirichlet boundary conditions\footnote{The
result cited is valid for bounded intervals of the real line; the
generalisation to semibounded intervals however is straightforwardly
obtained using their method.}. If we also denote by $G_0(x,x'|t)$
the full propagator in the position basis (corresponding to $e^{-i
\hat{H}t}$), we obtain
\begin{eqnarray}
\rho(t,t') &=& \frac{1}{4 M^2}\partial_x (\hat{C}_{t'}
\psi_0)^*(0)
\partial_x(\hat{C}_t \psi_0)(0)
G_0(0,0|t'-t) \\
d(t,N) &=& - \frac{1}{2M} \int_{- \infty}^0 dx (\hat{C}_T
\psi_0)^*(x)
\partial_x(\hat{C}_t \psi_0)(0)
G_0(x,0|T-t),
\end{eqnarray}
where $\hat{\rho}_0 = | \psi_0 \rangle \langle \psi_0 |$, with
$\psi_0$ having support in $(- \infty, 0]$.

Note that in the derivation of the equations above, the derivative
$\partial_x$ arises from the presence of a term $\hat{P}_+ \hat{H}
\hat{P}_-$ in the operator product in Eq. (\ref{densitydecf}). The
contribution of the potential $V(x)$ vanishes, and the only
contribution comes from the $\hat{p}^2$ of the kinetic energy.

For a free particle of mass $M$
\begin{eqnarray}
G_0(x,x',t) &=& \sqrt{\frac{M}{2 \pi i t}} e^{iM(x-x')^2/2t} \\
C_t(x,x') &=& \chi_-(x) \chi_-(x') \sqrt{\frac{M}{2 \pi i t}}
\left[e^{iM(x-x')^2/2t} - e^{iM(x+x')^2/2t} \right],
\end{eqnarray}
leading to
\begin{eqnarray}
\partial_x(\hat{C}_t \psi_0)(0) = \sqrt{\frac{M}{2 \pi i t}} \int_{-\infty}^0
dx \frac{-2 i M x}{t} e^{i \frac{M x^2}{2t}} \psi_0(x) := -2
\partial_x(\hat{U}_t \psi_0)(0),
\end{eqnarray}
where we assumed that $\hat{P}_- \psi_0 = \psi_0$.

It follows that
\begin{eqnarray}
\rho(t,t') = (\frac{\hat{p}}{M} \hat{U_t} \psi_0)(0)
(\frac{\hat{p}}{M} \hat{U_{t'}} \psi_0)^*(0) \sqrt{\frac{M}{2 \pi
i (t'-t)}} \label{rrrr}
\end{eqnarray}

\subsection{The classical limit}

To verify that the decoherence functional for the free particle
has the correct classical limit we consider a Gaussian initial
state centered around $x = - L$ and with mean momentum equal to
$p$
\begin{eqnarray}
\psi_0(x) = (\pi \sigma_0^2)^{-1/4} e^{-(x+L)^2/2 \sigma_0^2 + i p
x}. \label{psi0}
\end{eqnarray}
This state is localised within $[-\infty, 0 ]$ within an error of
order $e^{- L^2/\sigma^2}$. We then obtain
\begin{eqnarray}
\partial_x(\hat{C}_t \psi_0)(0) = \frac{2p}{\pi^{1/4}} \sqrt{\frac{\sigma_0}{\sigma^2(t)}}( \frac{t - t_{cl}} {M \sigma^2(t)} + i ) e^{-
(\frac{p^2}{2 m^2 \sigma^2(t)} ( t - t_{cl})^2 - i
\frac{p^2}{M}(t- t_{cl})}
\end{eqnarray}
where
\begin{eqnarray}
\sigma^2(t) = \sigma^2_0 (1 + i \frac{t}{M\sigma_0^2})
\end{eqnarray}
and  $t_{cl} = \frac{LM}{p}$ is the classical time-of-arrival.

Choosing $T$ very large ($T \rightarrow \infty$), so that the
classical time of arrival lies well within $[0, T]$ we see that
the bi-density $\rho(t, t')$ has a  singularity for $t = t'$ and
that it is sharply peaked in each of its arguments around $t_{cl}
= \frac{LM}{p}$, with a width $\delta$ of the order of
\begin{eqnarray}
\delta = \frac{M |\sigma(t_{cl})|}{p} = \frac{M
\sigma_0}{p}\sqrt{1 + \frac{L^2}{ \sigma_0^4 p^2}}. \label{delta}
\end{eqnarray}
Note that for large values of $L/p$ the width $\delta$ also becomes
very large. This is due to the fact that the free time evolution
causes the wave packet to spread in time. A very small value of
$\sigma_0$ (hence a large initial momentum uncertainty) leads to
large values of $\delta$.

Hence if we consider a coarse-grained history $\alpha_{cl}$ for the
time of crossing, which is centered around $t_{cl}$ and has a width
of $\Delta t
>> \delta$, and we denote as
$\alpha_{cl}'$ the complement of $\alpha_{cl}$, we obtain for the
conditioned decoherence functional
\begin{eqnarray}
|d_c(\alpha_{cl}, \alpha'_{cl})| = O (e^{-(\Delta t/\delta)^2}) \\
d_c(\alpha_{cl}, \alpha_{cl}) = 1 - O (e^{-(\Delta t/\delta)^2}),
\end{eqnarray}
hence we conclude that history $\alpha_{cl}$ will almost
definitely be realised, provided the particle crosses $x = 0$ at
some time within $[0, T]$.

Clearly there exists no classical limit, if either the initial
state is not sufficiently localised in position, or if it is too
localised so that the momentum uncertainty is very large, or if it
is a superposition of states with distinct value of momentum.

\subsection{Inclusion of measurement device}
In the consistent histories interpretation probabilities are only
defined, if the consistency condition is satisfied. For
time-of-arrival propositions this happens for  coarse-grained
histories like $\alpha_{cl}$  of Sec. 4.4, which essentially
correspond to the classical result. (Note that this result is
obtained after conditioning the decoherence functional upon
arrival). However, consistent histories refer to closed
systems--hence, to obtain a prediction for the time-of-arrival
probabilities in the general case we have to model the interaction
of the particle with the measuring device.

There are various models for such interaction with various degrees
of complexity \cite{model} (or for a more general case see
\cite{Schul}). We  consider here  a very simple one, which will
allow us to analysed some basic features of this procedure. We
model the pointer of the measuring device with a two level system.
The pointer is found in the state of lower energy, while the state
of higher energy corresponds to the detector having clicked. The
combined Hamiltonian of the system+apparatus is then
\begin{eqnarray}
\hat{H} =  \frac{\Omega}{2} (1 - \sigma_3) + \hat{H_0} + \epsilon
\hat{P}_+ \sigma_1, \label{HSA}
\end{eqnarray}
where $\hat{H}_0$ is the particle's Hamiltonian, and the
interaction characterised by a coupling constant $\epsilon$ is
switched on, only when the particle enters the region $ x > 0$. We
next construct histories similar to the ones of section 4, only
that they would refer to the properties of the pointer, i.e. they
would be constructed from projectors of the form
\begin{eqnarray}
\hat{E}_+ = \hat{1}_{particle} \otimes \left( \begin{array}{cc} 1
&0 \\ 0 & 0 \end{array} \right) \hspace{2cm} \hat{E}_- =
\hat{1}_{particle} \otimes \left( \begin{array}{cc} 0 &0 \\ 0 & 1
\end{array} \right),
\end{eqnarray}
inserted into the definition of histories of the form
(\ref{history}-\ref{history2}). Hence we seek the moment of
transition from the lowest energy state of the detector  to the
higher energy state.

We assume that the initial state is factorised $\hat{\rho} =
\hat{\rho}_0 \otimes \left( \begin{array}{cc} 0 &0 \\ 0 & 1
\end{array} \right) $, where $\hat{\rho}_0$ is the particle's density matrix. Using similar arguments we obtain the
following expressions for the densities defining the decoherence
functional
\begin{eqnarray}
\rho(t,t') = \epsilon ^2 Tr ( e^{-i \hat{H}_0(t'-t)}\hat{P}_+e^{-i
\hat{H}_0t}\hat{\rho}_0 e^{i \hat{H}_0t'} \hat{P}_+ ) +
O(\epsilon^4)
\\
d(t, N) = \epsilon Tr (e^{-i \hat{H}_0(T-t)}\hat{P}_+e^{-i
\hat{H}_0 t}\hat{\rho}_0 ),
\end{eqnarray}
while  $d(N,N) =1$. Again we can define the conditioned
decoherence functional on $[0, T]$ as $\rho_c(t,t') = \rho(t,t')/
d([0,T],  [0,T])$.

 This model does not solve the fundamental problem of defining a
probability density--one can easily check that the consistency
condition is approximately satisfied only for highly coarse-grained
histories centered around the classical time-of-arrival. On one
hand, this is not a surprise. Realistic measurement devices are much
more complex than the system described by the Hamiltonian
(\ref{HSA}). Moreover, they involve by necessity a degree of
irreversibility, which is incompatible with a unitary evolution law
(see for example a model in \cite{Hal98}): the states of detection
and no-detection are asymmetric, because the latter involves an
amplification procedure that leads to a macroscopic designation that
the particle has been detected. Such an asymmetry is incompatible
with unitary evolution.

On the other hand, the present model demonstrates a rather generic
feature associated with the measurement problem, that makes its
present felt in all interpretations of quantum theory that attempt
to describe the measurement process unitarily. In the consistent
histories approach measurement devices are thought of as quantum
systems, which are characterised by a consistent set of histories
for the pointer device, so that the values of the pointer can be
ascertained individually (and assigned probabilities) for a large
class of initial states of the measured system \cite{Omn8894,
Har93a}. This is  equivalent with obtaining a density matrix
diagonalised in a basis factorised with respect to the degrees of
freedom of the system and the measuring device, which  necessary, in
order to attribute definite values to the macroscopic pointer.

 In the model we presented here, this is clearly not the
case. There exist also very general theorems that state that such a
factorisation is in general not possible \cite{BG00}, see also the
discussion in \cite{Schl04, Adl01, Das05} and in a different but
related context the theorems of \cite{ BLM96}). The general argument
is rather simple and in the present context takes the following
form. The derivation of the decoherence functional for
time-of-arrival histories remains the same, whether the Hilbert
space is that of a single quantum system, or if it includes the
degrees of freedom of a measurement device or of an environment. The
non-zero value of the off-diagonal elements of the decoherence
functional would then still persist, except possibly for the case of
sufficient coarse-graining corresponding to the classical results.
Hence, it seems that, unless we introduce additional assumptions,
the consistent histories scheme cannot produce more detailed
information about the time-of-arrival probability, beyond the
determination of the classical limit.

The situation is different when the evolution of the particle
(rather than that of the measurement device) involves a degree of
irreversibility \cite{HaZa97}. Indeed, in the presence of a
decohering environment the evolution of the particle is closely
approximated (after a typically short decoherence time) by a
stochastic process, in which case one may employ the construction of
time-of-arrival probabilities sketched in section 2. However, the
time-of-arrival seems to depend rather crucially on properties of
the environment, which seems to destroy much "more" interference
than what is necessary to define a consistent set of
histories\footnote{There is a rather paradoxical situation in the
presence of environment. If the full quantum mechanical treatment of
system+environment is taken into account, in which case the full
evolution  law is unitary, we are faced with the quantum Zeno
effect. Any proposition about the system will be represented by
projection operators, and the operator $\hat{C}_t$ will still be a
degenerate unitary operator. If, however, one describes the effect
of the environment in terms of a non-unitary evolution (or a
stochastic process), an excellent approximation in many cases, no
such problem seems to arise. This suggests again that the quantum
Zeno effect is not robust.}

Another possibility would be to consider a different measuring
apparatus interacting with the particle at each moment of time.
This, however, would describe a different physical circumstance than
the one we consider in this paper. Here we consider the case that
the only information we can obtain from the system is a signal that
the particle has been detected at a specific moment of time, i.e.
there is a single event of measurement. Inserting multiple measuring
devices would be equivalent to letting the particle move within a
material that records its track (like a bubble chamber). In that
case our datum is a continuous history of the particle, and can be
treated within the theory of continuous time measurement. Such an
interaction inevitably causes the particle to decohere, and as such
its evolution can be well approximated by a stochastic process. The
derivation of the time of arrival would the be much simpler in this
case and would follow the general scheme of \cite{HaZa97}. This is,
however, different from the issue taken up in this paper.

\section{An operational description for time-of-arrival measurements}

The derivation of the decoherence functional for the time-of-arrival
propositions does not answer the question we asked at the beginning
of this paper: it is in principle possible to measure the
time-of-arrival in individual runs of an experiment, therefore
constructing the relative frequencies for a detection of the
particle within any time interval $[t_i, t_f]$. From these
frequencies of events we can construct in the limit of a large
number of runs a probability for the detection of the particle
within any time interval. One then is entitled to ask, how to obtain
this operationally meaningful probability density from the rules of
quantum theory.

The main contribution of the  histories formalism to the final
result is that it allows one to implement the continuous-time limit.
However, this passage refers not to a probability density, but to
the decoherence functional and one needs further assumptions in
order  to construct genuine probabilities. As the expression
(\ref{densitydecf}) for the decoherence functional demonstrates,
there is persistent interference between the different alternatives
for the time-of-arrival.

This problem is not specific to the time-of-arrival measurements. It
is a special case of a more general problem, that of defining a
probability density for the outcomes of measurement that take place
at more than one moment of time. This issue has been analysed in
\cite{Ana04, Ana05}. In sequential measurements it is possible to
obtain the probabilities in terms of a
Positive-Operator-Valued-Measure, whose mathematical form is
markedly similar to the "probabilities" constructed by the
consistent histories approach. We shall attempt to do the same here
for the time-of-arrival probabilities, thus exploiting the
convenient continuous-time limit incorporated in the decoherence
functional.

Note that the considerations in this section are purely
operational and employ the Copenhagen interpretation:  we do not
consider closed, individual systems, but are only interested about
probabilities obtained in specific measurement situations, which
we assume to refer to a statistical ensemble. Hence even though we
shall use the mathematical apparatus of consistent histories, the
context in which we work is markedly distinct.

\subsection{Probabilities for sequential measurements}
\paragraph{Discrete spectrum}
Let us consider the two-time measurement of an observable $\hat{x}
= \sum_i \lambda_i \hat{P}_i$ with discrete spectrum. Writing
$\hat{Q}_i = e^{i \hat{H}t} \hat{P}_i e^{-i \hat{H}t}$, we
construct the probabilities for the most-fine grained two-time
results
\begin{eqnarray}
p(i,0; j,t) = Tr (\hat{Q}_j \hat{P}_i \hat{\rho}_0 \hat{P}_i) =
|\langle|\hat{\rho}_0|i \rangle|^2 |\langle i|e^{-i \hat{H}t}| j
\rangle|^2 \label{finegrained}
\end{eqnarray}

Irrespective of the interpretation of the measurement process, the
probabilities (\ref{finegrained}) refer to the most elementary
alternatives that can be unambiguously determined in the
experimental set-up corresponding to the sequential measurement of
$\hat{x}$. Therefore, they can be  employed to construct
probabilities for general sample sets $U_1$, $U_2$ on the spectrum
$\Omega$ of $\hat{x}$, namely
\begin{eqnarray}
p(U_1,0; U_2, t) = \sum_{i \in U_1} \sum_{j \in U_2} p(i,0; j,t).
\label{POVM2}
\end{eqnarray}
The total probability is normalised
\begin{eqnarray}
p(\Omega,0; \Omega,t) = \sum_{ij}Tr (\hat{Q}_j \hat{P}_i
\hat{\rho}_0 \hat{P}_i) = 1,
\end{eqnarray}

a property that follows from the fact that $\sum_i \hat{P}_i =
\sum_j \hat{Q}_j = \hat{1}$.

 Eq. (\ref{POVM2}) defines a POVM for two-time
measurements. Note the difference from equation (\ref{decfundef}).
Equation (\ref{decfundef}) is valid only for the most fine-grained
alternatives. Any further coarse-graining is done by summing only
the elementary probabilities that correspond to the value of the
decoherence functional for the most fine-grained histories. This
result implies that we can use the decoherence functional to
construct POVMs for sequential measurements and may expect to repeat
do the same for the case of time-of-arrival.

\paragraph{Continuous spectrum}
When we consider the case of an operator with discrete spectrum a
problem appears. There are no fine-grained projectors and the
choice of the elementary quantum probabilities, from which one may
build the general probabilities for measurement outcomes cannot be
made uniquely.

The immediate generalisation of Eq. (\ref{finegrained}) for the
measurement of an operator with a continuous spectrum is
\begin{eqnarray}
p(x_1,0; x_2,t) =  |\langle x_1 |\hat{\rho}_0|x_1 \rangle|^2
|\langle x_1|e^{-i \hat{H}t}| x_2 \rangle|^2. \label{ff}
\end{eqnarray}
However, this does not define a proper probability density, because
it is not normalised to unity
\begin{eqnarray}
\int dx_1 \int dx_2 \, p(x_1,0; x_2,t) = \infty.
\end{eqnarray}
This is due to the fact that there can be no measurements of
infinite accuracy. One has, therefore, to take into account the
finite width of any position measurement, say $\delta$. This
quantity depends on the properties of the measuring device--for
example the type of the material that records the particle's
position.

 The simplest procedure
is to consider the measurement of a self-adjoint operator
$\hat{x}_{\delta} = \sum_i x_i \hat{P}^{\delta}_i$, where
$\hat{P}^{\delta}_i$ is a projection operator corresponding to the
interval $[x_i - \frac{\delta}{2}, x_i + \frac{\delta}{2}]$. In
that case we may immediately construct the fine-grained
probabilities
\begin{eqnarray}
p_{\delta}(i,0; j,t) = Tr (\hat{Q}^{\delta}_j \hat{P}^{\delta}_i
\hat{\rho}_0 \hat{P}^{\delta}_i), \label{fine}
\end{eqnarray}
from which we may construct probabilities for general sample sets
$U_1$ and $U_2$:
\begin{eqnarray}
p_{\delta}(U_1,0;U_2,t) = \sum_{i \in U_1} \sum_{j \in U_2} Tr
(\hat{Q}^{\delta}_j \hat{P}^{\delta}_i \hat{\rho}_0
\hat{P}^{\delta}_i).
\end{eqnarray}
 Strictly speaking one may only consider sample sets that are
unions of the elementary sets that define our lattice. If,
however, the size of the sample sets $L$ is much larger than
$\delta$, we may approximate the summation with an integral. This
amounts to defining the continuous version of probabilities
(\ref{fine})
\begin{eqnarray}
p_{\delta}(x_1,t_1;x_2, t_2) = \hspace{4cm}\\ \nonumber Tr \left(
e^{i\hat{H}(t_2-t_1)} \hat{P}^{\delta}_{x_2}
e^{-i\hat{H}(t_2-t_1)} \hat{P}^{\delta}_{x_1} \hat{\rho}(t_1)
\hat{P}^{\delta}_{x_1} \right),
\end{eqnarray}
where we denoted $\hat{P}_x^{\delta} =
\int_{x-\delta/2}^{x+\delta/2} dy |y \rangle \langle y|.$

The important result of this analysis is that the probabilities
for sequential measurements (\ref{fine}) depend strongly on the
resolution $\delta$ of the measuring device. This dependence is
very strong: even probabilities of sample sets coarse-grained at a
scale much larger than $\delta$ exhibit a very strong dependence
on $\delta$. From a mathematical point of view this dependence is
a consequence of the fact that the off-diagonal elements of the
decoherence functional between fine-grained multi-time measurement
outcomes do not vanish and are generically of the order of
magnitude of the probabilities themselves. Hence, when we compare
a probability corresponding to a value $\delta$, with another one
corresponding to $2 \delta$, they differ by an amount proportional
to the corresponding off-diagonal terms of the decoherence
functional, which is in general substantially large.
 The reader is referred to \cite{Ana05} for extensive
discussion and generalisations.

The construction probabilities for the outcomes of sequential
measurements consists of two steps. First we identify the most
fine-grained alternatives compatible with the measuring device at
hand and we construct the corresponding elementary quantum
probabilities by using the rule (\ref{decfundef}). These fine-
grained alternatives (referred to by the index $a$) correspond to
specific functions $F_a[x(\cdot)]$ on the space of paths
$\Omega^I$. The elementary probabilities will be
\begin{eqnarray}
p(a) = d(F_a, F_a).
\end{eqnarray}
In the equation above the decoherence functional is viewed as a
bi-linear functional on $\Omega^I$. For example, in the two-time
measurement of position the  functions take the form
$F_{ij}[x(\cdot)] = \chi_i^{\delta}(X_{t_1}
\chi_j^{\delta}(X_{t_2})$, and
\begin{eqnarray}
p_{\delta}(i,0; j,t) = d(F_{ij}, F_{ij}), \label{ee}
\end{eqnarray}

The next step involves the summation over those elementary
probabilities to construct an additive measure that assigns
probabilities to every sample set obtained by the coarse-graining
of the elementary alternatives. We shall apply this procedure to
the construction of time-of-arrival probabilities.

\subsection{ POVMs for time-of-arrival probabilities}

Our contention is that the analysis of sequential measurements above
may be transferred to the case of time-of-arrival measurements,
because they share the crucial feature that they do not refer to the
properties of a physical system at a single moment of time. This
implies that the decoherence functional (\ref{densitydecf}) may be
employed for the  construction of  a POVM on $[0, T]$ for the time
of arrival probabilities, in analogy to that of (\ref{fine}).

The diagonal elements $\rho(t,t) \delta t^2$ of the decoherence
functional  (\ref{densitydecf}) is essentially the modulus square of
the amplitude that is obtained by the sum over all paths that cross
the surface $x = 0$ within the interval $t + \delta t$. While the
amplitude is obtained unambiguously through path integrals, its
square cannot define a proper probability density, because of the
presence of a term $\delta t^2$ rather than a $\delta t$ one, but
also because $\rho(t,t)$ diverges\footnote{It is interesting to note
that the $\delta t^2$ dependence disappears if the decoherence
condition for histories holds--see \cite{Yam96}.}. This divergence
is analogous to that of (\ref{ff}) for sequential measurements of
position, and can be removed in a similar manner by assuming a
finite temporal resolution $\tau$.
 Hence, we consider elementary intervals
$[t_i, t_{i+1}]$ of width $\tau$, $\tau$ corresponding to the
temporal resolution of our measurement device. The elementary
probabilities will be
\begin{eqnarray}
\rho^{\tau}_i = \int_0^T dt \int_0^T dt' \rho(t,t')
\chi_i^{\tau}(t) \chi_i^{\tau}(t'), \label{disctime}
\end{eqnarray}
where $\chi_i^{\tau}$ is the characteristic function of the set
$[t_i, t_{i+1}]$. One then may employ these probabilities to
construct any probability corresponding to a set $U$ constructed
from the elementary cells $[t_i, t_{i+1}]$. By definition $\sum_i
\chi_i^{\tau} = \chi_{[0, T]}$, hence the set of all $i$ together
with the event of no detection form a proper resolution of the
unity.

The reader may object at this point that this leads us back to the
discrete-time expression for the diagonal elements of the
decoherence functional. This is not the case, because the
probabilities (\ref{disctime}) involve the sum over {\em all}
continuous paths that are detected in the time interval $[t_i,
t_{i+1}]$--hence it involves the contribution of any discretisation
within $[t_i, t_{i+1}]$.

It is  more convenient to avoid the discretisation procedure and
construct a POVM on the continuous sample space $[0, T]$--see
(\cite{Ana05}) for the analogous procedure in sequential
measurements. For this purpose we introduce a family of smeared
delta functions $f_{\tau}(s, s')$ characterised by the parameter
$\tau$, which satisfy the following properties
\begin{eqnarray}
\int_0^T ds f^{\tau}(s, s') = \chi_{[0, T]}(s').
\end{eqnarray}
One may consider for example the following functions
\begin{eqnarray}
f_{\tau}(s, s') = \frac{1}{T} \sum_{n = - [T/\tau]}^{[T/\tau]}
e^{i \frac{n \pi}{T}(s - s')}.
\end{eqnarray}
For practical purposes these are well approximated by the
Gaussians (as long as $T >> \tau$)
\begin{eqnarray}
f^{\tau}(s,s') = \frac{1}{\sqrt{2 \pi} \tau} e^{ - \frac{(s -
s')^2}{2 \tau^2}}. \label{Gauss}
\end{eqnarray}

Thus we may define the elementary probabilities in analogy to
(\ref{fine}) as
\begin{eqnarray}
p^{\tau}(t) = \int ds ds' \sqrt{f^{\tau}(t,s)} \sqrt{f^{\tau}(t,s')}
\rho(s, s'), \label{ppp}
\end{eqnarray}
and construct from them the probabilities for any set $U \subset
[0, T]$ as
\begin{eqnarray}
p^{\tau}(U) = \int_U dt p^{\tau}(t).
\end{eqnarray}

Up to an error of order $\tau$ this is equivalent with the
probabilities obtained by coarse-graining the elementary
discrete-time probabilities (\ref{disctime})\footnote{The use of the
square root in (\ref{ppp}) is necessary in order to guarantee the
proper dimensions of the probability density (dimensions of
$[T]^{-1}$). Another way to see this is by noticing that
$\sqrt{\sqrt{2 \pi}\tau f^{\tau}}(t)$, for the Gaussian
(\ref{Gauss}) is a smeared characteristic function for the interval
$[t-\sqrt{2 \pi} \tau, t+ \sqrt{2 \pi}\tau]$, thus corresponding to
a smeared version of (\ref{disctime}), which needs to be divided by
$\sqrt{2 \pi} \tau$ in order to define a probability density. Note
also that if the decoherence condition holds, Eq. (\ref{ppp})
becomes, as it must, the Gaussian smearing of the probability
distribution for arrival times.}.

 In effect, we associate to each set $U$ the positive
operator
\begin{eqnarray}
\hat{\Pi}(U) = \int_U dt \hat{R}_t \hat{R}^{\dagger}_t,
\label{POVMt}
\end{eqnarray}
where
\begin{eqnarray}
\hat{R}_t = \int ds \, \sqrt{f^{\tau}(t,s)} \hat{C}_s \hat{P}_-
\hat{H} \hat{P}_+ e^{i \hat{H}s}, \label{Rt}
\end{eqnarray}
is an operator that corresponds to the sum over all paths that lie
within $[-\infty, 0]$ and cross through to $[0, \infty)$ within a
time interval of width $\tau$ around $t$.

These positive operators do not yet define a POVM, because the
corresponding probabilities do not add-up to unity. We have to
also include the event $N$ of no detection. The normalisation
condition implies that a positive operator $\hat{\Pi}(N)$ should
be defined as
\begin{eqnarray}
\hat{\Pi}(N) = \hat{1} - \int_0^T dt \hat{R}_t
\hat{R}^{\dagger}_t.
\end{eqnarray}

The operator $\hat{\Pi}(N)$ is indeed positive, because
\begin{eqnarray}
 \int_0^T dt p^{\tau}(t) \leq \sup_{s,s'  \in [0, T]} (\int_0^T dt
\sqrt{f^{\tau}(t,s)} \sqrt{f^{\tau}(t,s')} )\int_0^T ds \, ds' \,
\rho(s,s').
\end{eqnarray}
Since $f^{\tau}$ is a smeared delta function, the term $\int_0^T dt
\sqrt{f^{\tau}(t,s)} \sqrt{f^{\tau}(t,s')}$ is maximised for $s
=s'$, in which case it equals $\int_0^T dt f^{\tau}(t,s) = \chi_{[0,
T]}(s) \leq 1$. Hence $\int_0^T dt p^{\tau}(t) \leq \int_0^T ds ds'
\rho(s,s') = d([0, T], [0, T]) \leq 1$. Hence
\begin{eqnarray}
p(N) = Tr (\hat{\rho} \hat{\Pi}(N)) = 1 - \int_0^T dt \, p^{\tau}(t)
\geq 0,
\end{eqnarray}
for all $\hat{\rho}$.

 We have thus constructed a POVM for the
time-of-arrival essentially by summing over all possible paths that
correspond to crossing the $x = 0$ surface within a time interval of
width $\tau$; $\tau$ was essentially introduced as a
"regularisation" parameter. In general, the POVM is expected to
depend strongly upon its value. The key idea employed in this
derivation is that time-of-arrival probabilities are not
fundamentally different from the probabilities that correspond to
measurements that take place at more than one moment of time.
Whenever the measured quantity is continuous, it is necessary to
introduce a parameter that determines the resolution of the
measuring device, and it turns out that the constructed
probabilities depend strongly on this parameter. Measurements of the
time-of-arrival like sequential measurements of position seem to be
strongly contextual, namely to depend strongly on the specific
measurement device employed in their determination \cite{Ana05}.

Our derivation relied on two assumptions. The first one is  that
the reduction rule may be employed consistently for the
incorporation of {\em any information} we may obtain about a
quantum system (and not only for the results of actual
measurements as in section 3.2). The second assumption is that the
construction of probabilities for sequential measurements may be
applied in the context of time-of-arrival measurements through a
generalisation of Eq. (\ref{ee}). The key mathematical input
arises from the histories description, namely the fact that it is
possible to construct a sample space for the values of the
time-of-arrival by considering continuous-time-histories of the
system.

\subsection{An explicit calculation: the free particle}

We shall now compute the POVM (\ref{POVMt}) explicitly for the case
of a free particle. This case is particularly interesting, because
it allows the comparison with a well-established result, namely the
POVM constructed by Kijowski \cite{Kij74}. Kijowski's POVM for the
time-of-arrival of a free particle assigns to any pure state
$\psi_0$  a probability density $p(t, \psi_0)$, which is normalised
to unity in the interval $(-\infty, \infty)$,

\begin{eqnarray}
p(t, \psi_0) = |\int_0^{\infty} dp \left(\frac{p}{ 2\pi m}
\right)^{1/2} e^{-ip^2t/2M} \psi(p)|^2  \nonumber \\
+ |\int^0_{-\infty} dp \left(\frac{-p}{ 2\pi m} \right)^{1/2}
e^{-ip^2t/2M} \psi(p)|^2 \label{Kij}.
\end{eqnarray}

To construct the POVM (\ref{POVMt}) for a free particle we use Eq.
(\ref{rrrr}) for the decoherence functional. Since the integration
in (\ref{ppp}) involves the square roots of the smeared
delta-functions, which have a width of order $\tau$, we may within
an error of order $O(\tau/T)$ substitute the range of integration
 $\int_0^T ds \int_0^T ds' \rightarrow
 \int_{-\infty}^{\infty} ds \int_{-\infty}^{\infty} ds'$ and employ
 the Gaussian smearing functions (\ref{Gauss}).

 The probability density associated to (\ref{POVMt}) can be written in
 the momentum representation as follows
 \begin{eqnarray}
p(t) = \frac{1}{2 \pi} \int dp \int dp' \frac{p \, p'}{M^2}
R(p,p',t) \tilde{\psi}_0(p) \tilde{\psi}_0^*(p'),
 \end{eqnarray}
where $\tilde{\psi}_0$ is the Fourier transform of $\psi_0$ and
\begin{eqnarray}
R(p,p',t) =  \int_{-\infty}^{\infty} ds \int_{-\infty}^{\infty} ds'
\; \sqrt{f^{\tau}(t,s)} \sqrt{f^{\tau}(t,s')} e^{- i \frac{p^2}{2M}
s + i \frac{p'^2}{2M}s'} \sqrt{\frac{M}{2 \pi i (s-s')}}. \label{Rf}
\end{eqnarray}

Changing variables to $u = \frac{1}{2} (s+s')$ and $v = s-s'$, we
note that
\begin{eqnarray}
\sqrt{f^{\tau}(t,s)} \sqrt{f^{\tau}(t,s')} = f_{\tau} (u - t) e^{-
\frac{v^2}{8 \tau^2}}.
\end{eqnarray}

Within an error of order $O(\tau/T)$, the function  $f_{\tau} (u -
t)$ behaves as a delta-function when integrated over $u$, thus
leading to
\begin{eqnarray}
p(t) \simeq \frac{1}{2 \pi} \int dp \int dp' \frac{p \, p'}{M^2}
e^{-i (\frac{p^2}{2M} - \frac{p^2}{2M})t} \, r
\left(\frac{E_p+E_{p'}}{2} \right)\tilde{\psi}_0(p)
\tilde{\psi}_0^*(p'), \label{pr}
\end{eqnarray}
where $E_p = \frac{p^2}{2M}$ and
\begin{eqnarray}
r(\epsilon) &=& \sqrt{\frac{M}{2 \pi}} \int_{-\infty}^{\infty} dv
\frac{e^{- \frac{v^2}{8 \tau^2}-i \epsilon v}}{\sqrt{iv}}
\nonumber
\\
&=& \sqrt{\frac{2M \tau}{ \pi}} \int_0^{\infty}dy \;  \frac{ e^{ -
\frac{y^2}{2}} [ \cos (2 \epsilon \tau y) + \sin(2 \epsilon \tau
y)]}{\sqrt{y}}. \label{re}
\end{eqnarray}

The integral in Eq. (\ref{re}) can be computed explicitly in terms
of modified Bessel functions; however the physically interesting
information is found in specific regimes for which $r(\epsilon)$
takes a simple form. For $\epsilon \tau <<1$, the leading
contribution to the integral is a constant, leading to
\begin{eqnarray}
r(\epsilon) = \frac{\Gamma(\frac{1}{4})}{2^{3/4}}\sqrt{\frac{M
\tau}{2 \pi}},
\end{eqnarray}
which implies that the probability density $p(t)$ is proportional
to $\tau^{1/2}$, when $\tau \rightarrow 0$. Hence, at the limit
that the temporal resolution tends to zero the particle is never
detected  crossing the surface $x = 0$.

 The physically interesting regime
 corresponds to $\epsilon \tau >>1$. The parameter $\epsilon$  appears in
(\ref{pr}) as the particle's energy, while  $\tau$ is  the temporal
resolution of the measurement device. According to a common
interpretation of the time-energy uncertainty principle, $\tau$
cannot be smaller than $(\Delta E)^{-1}$, where $\Delta E$ is the
energy spread of the wave-functions. Hence for any wave-function
with relatively small energy spread ($\Delta E/ E << 1$), one
expects that $\tau E >> 1$. In general, it suffices that the
wave-function has support only for values of energy much larger than
$\frac{1}{\tau}$. The resolution $\tau $ is by assumption much
smaller than $t_{cl}$, hence this range of energies is well defined,
whenever $E t_{cl}
>> 1$. Since $t_{cl} = \frac{ML}{\bar{p}}$, where $\bar{p}$ is the
initial state's mean momentum,
 the condition above is equivalent
to $\bar{p} L >> \hbar$, which is always satisfied in any
macroscopic configuration for the measurement of the
time-of-arrival.

At the limit $\epsilon \tau >> 1$,
\begin{eqnarray}
\int_0^{\infty}dy \;  \frac{ e^{ - \frac{y^2}{2}} [ \cos (2
\epsilon \tau y) + \sin(2 \epsilon \tau y)]}{\sqrt{y}} \simeq
\sqrt{\frac{\pi}{\epsilon \tau}},
\end{eqnarray}
hence the dependence on $\tau$ drops from the  probability density
(\ref{pr})
\begin{eqnarray}
p(t) \simeq \frac{1}{\pi} \int dp \int dp' \frac{p \, p'}{M
\sqrt{\frac{1}{2} (p^2 +p'^2)}}  e^{- i(\frac{p^2}{2M} -
\frac{p^2}{2M})t} \, \tilde{\psi}_0(p) \tilde{\psi}_0^*(p'),
\label{pfin}
\end{eqnarray}

The POVM (\ref{pfin}) is defined for positive values of time, and
for wave functions that satisfy $\hat{P}_+ |\psi_0 \rangle = 0$. To
compare (\ref{pfin}) with Kijowski's POVM of Eq. (\ref{Kij}), which
is normalised to unity by integration over for all times $t \in
(-\infty, \infty)$, it is convenient to also extend the domain of
the probability distribution (\ref{pfin}) to the whole real axis for
time, by requiring that the extended POVM is invariant under the
combined action of the parity and time-reversal transformations
\cite{Kij74}. We employ the convention that the negative times
correspond to the crossing of $x = 0$ from the right, i.e. to
initial states that have support on values of position $x \in [0,
\infty)$. We then construct an equal-weight convex combination of
the probability distribution (\ref{pfin}) for positive $t$ with its
counterpart for negative $t$. We thus obtain the extension of the
probability distribution (\ref{pfin}) for all real values of time
and all initial wave-functions $\tilde{\psi}_0(p)$
\begin{eqnarray}
p_{ext}(t) \simeq \frac{1}{2\pi} \int dp \int dp' \frac{p \, p'}{M
\sqrt{\frac{1}{2} (p^2 +p'^2)}}  e^{- i(\frac{p^2}{2M} -
\frac{p^2}{2M})t} \, \tilde{\psi}_0(p) \tilde{\psi}_0^*(p').
\label{pfin2}
\end{eqnarray}

If the wave-function is sharply concentrated around the mean value
$\bar{p}$, i.e. if $\Delta p << |\bar{p}|$, then  $p^2 + p'^2 = 2 p
p' + O((\Delta p/ \bar{p})^2)$ within the integration in
(\ref{pfin}). The probability density (\ref{pfin2}) is then
identical with (\ref{Kij}). Integrating $p_{ext}(t)$ over $t \in
(-\infty, \infty)$ we obtain

\begin{eqnarray}
 \int_{-\infty}^{\infty} dt \, p_{ext}(t) = 1 - \int dp \;
\tilde{\psi}^*_0(-p) \tilde{\psi}_0(p).
\end{eqnarray}

We see therefore that  $p_{ext}(t)$ is normalised to unity, if the
state $\tilde{\psi_0}(p)$ has support only on positive (or only on
negative) values of momentum. In this case, $\hat{\Pi}(N) = 0$, i.e.
all particles are eventually detected. In general, a non-zero
probability $p(N)$ of non-detection is due to the fraction of
particles in the statistical ensemble, which move away from the
crossing surface $x = 0$.

 We conclude therefore that the POVM (\ref{POVMt}) leads to the same
 probability distributions for the time-of-arrival with Kijowski's POVM (\ref{Kij})
  for all initial wave-functions that (i) have support in
values of momentum $|p| >> \sqrt{\frac{2m}{\tau}}$ and (ii) satisfy
$\Delta p << |\bar{p}|$. This regime includes the classical limit
(e.g. wave-functions of the form (\ref{psi0}) with $\sigma_0 p
>>1$ ), but also states that are not sharply localised in
position, like superpositions of macroscopically distinct
wave-packets, or even superpositions of states with different
value of momentum--as long as $\Delta p$ remains much smaller than
$p$. Outside this regime, the POVMs (\ref{POVMt}) and (\ref{Kij})
provide different predictions.

To make the points above more explicit we consider an initial
Gaussian state
\begin{equation}
\tilde{\psi}_0(p) = (a^2/2\pi)^{1/4} \exp \left(- \frac{a^2}{4} (p -
\bar{p})^2 + i p L \right) \label{ggg}
\end{equation}
 of mean position $-L$ and mean momentum
$\bar{p}$. The momentum spreads equals to $a$. In leading order to
$\frac{1}{a \bar{p}}$ the probability distributions for the
time-of-arrival predicted by the POVMs (\ref{pfin}) and (\ref{Kij})
coincide
\begin{eqnarray}
p(t) \simeq \frac{\bar{p}}{M a\sqrt{8\pi}} e^{ - \frac{2}{a^2}(L -
\bar{p} t/M)^2}.
\end{eqnarray}
The difference between the two POVMs is of order $1/(\bar{p}a)^2$
and even for a relatively large value $\frac{1}{\bar{p}a} \simeq
0.1$ as in Fig. 1 their graphs are practically indistinguishable.

\begin{figure}[h]
\includegraphics[height=5cm]{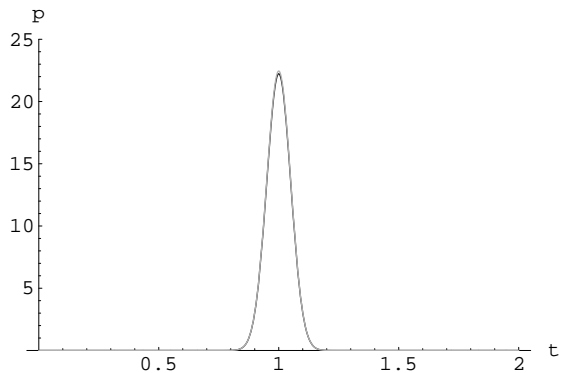}
\caption{ \small The probability distribution of the time-of-arrival
for a Gaussian initial state (\ref{ggg}) of mean position $-L$ and
mean momentum $\bar{p}$. It is sharply peaked around the value
$t_{cl} = M L / \bar{p}$. In fact, this plot contains the
probability distributions provided by both POVMs (\ref{pfin}) and
(\ref{Kij}), but even for the relatively large value   of $\Delta p
= 0.1 \bar{p}$ we employed here, the two curves almost coincide.}
\end{figure}

We next consider an initial state, which is a superposition of two
Gaussians with the same mean position $-L$, but different mean
momenta $\bar{p}_1$ and $\bar{p}_2$
\begin{eqnarray}
\tilde{\psi}_0(p) = \left(\frac{a^2}{4\pi} \right)^{1/4} \left[ e^{-
\frac{a^2}{4} (p - \bar{p}_1)^2 + i p L} + e^{- \frac{a^2}{4} (p -
\bar{p}_2)^2 + i p L } \right].
\end{eqnarray}
For $a \bar{p}_1 >> 1$, $a \bar{p}_2 >> 1$ and $a |\bar{p}_1 -
\bar{p}_2| >> 1$ the leading contribution to the probability
distribution obtained by the POVM (\ref{Kij}) is

\begin{eqnarray}
p(t) \simeq \frac{1}{M a\sqrt{8\pi}} \left[ \frac{\bar{p}_1}{2} e^{
- \frac{2}{a^2} (L - \bar{p}_1 t/M)^2} + \frac{\bar{p}_2}{2} e^{ -
\frac{2}{a^2}(L - \bar{p}_1 t/M)^2} \right. \nonumber
\\ \left. + \sqrt{\bar{p}_1 \bar{p}_2} \; e^{ - \frac{2}{a^2}(L -
\frac{1}{2}(\bar{p}_1 + \bar{p}_2) t/M)^2}
\cos\left(\frac{\bar{p}_1^2t}{2M} - \frac{\bar{p}_2^2t}{2M} \right)
\right], \label{222}
\end{eqnarray}
and the leading contribution to the probability contribution
obtained by the POVM (\ref{pfin}) is
\begin{eqnarray}
p(t) \simeq \frac{1}{M a\sqrt{8\pi}} \left[ \frac{\bar{p}_1}{2} e^{
- \frac{2}{a^2} (L - \bar{p}_1 t/M)^2} + \frac{\bar{p}_2}{2}  e^{ -
\frac{2}{a^2}(L - \bar{p}_1 t/M)^2} \right. \nonumber
\\ \left. + \frac{\bar{p}_1 \bar{p_2}}
{\sqrt{\frac{1}{2}(\bar{p}_1^2 + \bar{p}_2^2)}} \; e^{ -
\frac{2}{a^2}(L - \frac{1}{2}(\bar{p}_1 + \bar{p}_2) t/M)^2}
\cos\left(\frac{\bar{p}_1^2t}{2M} - \frac{\bar{p}_2^2t}{2M} \right)
\right]. \label{333}
\end{eqnarray}

We see that near the two classical values of the time of arrival
that correspond to each of the two wavepackets, the two probability
distributions coincide. However, for intermediate values of time,
they differ. They both manifest an oscillatory behaviour there,
which is characteristic of interference between the two classical
values of the time-of-arrival.
 From Eqs. (\ref{222}) and (\ref{333}) we readily see that the oscillation amplitudes are different
in this regime and their ratio is given  by the quantity
$\frac{1}{2}\left(\frac{\bar{p}_1}{\bar{p}_2} +
\frac{\bar{p}_2}{\bar{p}_1} \right)$. When this  becomes appreciably
larger than unity, i.e. if the difference in mean momentum between
the two Gaussians becomes comparable with the mean momenta
themselves, the behaviour of the two distributions in the
oscillatory region becomes substantially different.

 It is important to emphasise that the domain of
applicability of the
 POVM (\ref{POVMt}) is much larger than that of the free particle
 case (for a
generalisation
 of Kijowski's distribution see also \cite{BSPME00}). It can be in
 principle applied for systems described by arbitrary
 Hamiltonians. Moreover, it is constructed through a general
 argumentation that does not refer only to the time-of-arrival
 and it can be easily generalised to  systems with finite-dimensional Hilbert
 spaces, for which there is no analogue of (\ref{Kij}).
To see this one may consider equations (\ref{POVMt}) and (\ref{Rt}).
The only objects appearing in the definition of the POVM is the
Hamiltonian (together with its propagator $\hat{C}_t$) and the
projection operators corresponding to the two alternatives of
detection. The POVM (\ref{POVMt}) is therefore completely general.
It can be applied for example the description of the particle being
coupled to a measuring device, in which case the alternatives will
correspond to projectors of the device's Hilbert space, and the
Hamiltonian will contain an interaction term between particle and
measurement device. It can also accommodate interactions with the
environment--i.e. further terms in the Hamiltonian. Its more
immediate application would be the study of tunneling probabilities.
This POVM may also refer to systems other than particles (e.g.
multi-level atoms). Its derivation is only based on properties of
Hilbert space operators and for this reason it can be applied to any
physical context.

We chose to elaborate on the free particle case and ignore the
effects of any measuring devices. The reason is that this system
contains no other parameters other than the particle's mass (no
couplings) and for this reason the only relevant time scale is
$t_{cl}$. This allowed us to identify a
 physically relevant regime in which the time-of-arrival
 probabilities do not depend strongly on $\tau$.  Thus we were able to
 compare our result with Kijowski's POVM. This, however,
 cannot be expected to hold for  general systems, which may involve
 time-scales of the same order of magnitude or smaller than
 $\tau$. In the general case, the POVM (\ref{POVMt}) is expected to
 have a
 strong dependence on $\tau$ even in  physically relevant
 regimes.

\subsection{The problem of contextuality}

 We saw that in
the free particle case, there exists a regime, in which the
time-of-arrival probabilities are rather insensitive to $\tau$, but
this simple behaviour cannot be expected to hold for systems with
more complicated Hamiltonians that involve additional time-scales.
The POVM (\ref{POVMt}) will in general be strongly dependent on
$\tau$, and hence the probabilities for time will be strongly
dependent upon the measurement scheme employed for their
determination.

 This contextuality of time-measurements in quantum theory
has been emphasised by Landauer in his study of tunneling times
\cite{Land} (see also a related discussion with reference to the
quantum Zeno effect \cite{MiSu77}). However, this result is not a
consequence of time being a special or distinguished variable. This
type of contextuality is generic in quantum theory, once we consider
measurements that do not refer to a single moment of time--e.g.
sequential measurements of a continuous variable. This is a
necessary consequence of the formalism of quantum theory: the
evolution of the quantum state involves a linear law, while
probabilities are quadratic with respect to the state. Hence, it is
(in general) impossible to construct a natural probability measure
for the outcomes of any measurement that reveals information that
pertain to a system's dynamics (sequential, time-of-arrival,
continuous measurements). This problem can be seen from different
angles: from the fact that the dependence on time of the quantum
probabilities do not define a probability measure and hence the
continuous-time limit is not well defined (as in section 5.2); from
the fact that the natural measure for histories (\ref{decfundef}) is
non-additive; from the necessity to regularise the path integral
amplitudes for the time-of-arrival in order to define probabilities;
from the fact that there is interference between different
alternatives for the value of the time-of-arrival. One therefore has
to introduce an additional structure (external to the physical
parameters of the system). The simplest such structure is the
specification of the most fine-grained outcomes that can be recorded
by  a measurement device. In the case of observables with discrete
spectrum, this is provided naturally by the formalism. For
continuous variables, however, it is not, and this brings inevitably
the introduction of a scale for the fine-grained alternatives.

We cannot evade this problem by enlarging our system, including for
example a quantum measuring device or an environment. The problem is
a consequence of the interplay between the quantum probability rule
and the unitary dynamics. It will appear in any closed system (and
will be accompanied by the quantum Zeno effect). Indeed, our
arguments here were very general and hold with few modifications for
an arbitrary Hilbert space and with reference to the detection of
any quantum event. To avoid this problem (which can take a rather
extreme form \cite{Ana05}) we have to abandon either the probability
rule or the dynamics, and neither one of these steps is easy to
take.

On the other hand, the acceptance of this contextuality is very
disturbing. The devices that determine the time of arrival are not
different in nature from the ones that measure a particle's
position. (This is reflected in the fact that the histories for the
time-of-arrival are written in terms of spectral elements of the
position operator.) The only distinctive character in the
time-of-arrival measurements is that the "observable" quantity is
the reading of the clock, which is associated to the time of
detection. In position measurements, however, the fuzziness due to
the finite resolution of the measurement device is relatively small,
when the sampling of the measurement results are sufficiently
coarse-grained. On physical grounds one would expect that
coarse-graining at a scale much larger than the temporal resolution
of the measurement device would give results independent of the
device. Unlike the case of the free particle there is no reason to
expect this for a general Hamiltonian--unless one considers the
highly coarse-grained samplings around the classical equations of
motion.

\section{Conclusions}

We have considered the problem of constructing a probability
density for the time-of-arrival. Our main guideline was the fact
that time appears as an external parameter in quantum theory. We
relied on the histories formalism, because they allow the natural
definition of probabilities about the time-of-arrival.

In our perspective, the most severe problem in the determination of
the time-of-arrival probability is the fact that the quantum states
do not correspond to densities with respect to time. For this reason
it is very difficult to obtain the continuous-time limit in a
natural way. The naive way of taking the continuous-time limit gives
very bad results, as it is plagued by the quantum Zeno effect. The
first alternative we tried is to employ a more strict operational
interpretation of the wave packet reduction, namely that it can only
be applied as a result of a system's {\em physical}  interaction
with a measurement device, and not when we obtain information about
the system through inference from the lack of a detection signal.
Again, the continuous-time limit was pathological and involved the
introduction of an arbitrary temporal resolution.

 We then considered this problem in the light of the consistent
 histories approach. This suggests that the continuous-time limit
 should be taken at the level of amplitudes and not of
 probabilities, and for this reason it can be taken unambiguously.
 The consistent histories framework, however, is not sufficient
 for the definition of probabilities--there is `interference' between different
 values of the time-of-arrival. This problem is
 aggravated by the presence of the quantum Zeno effect.

Nonetheless, the mathematical benefits conveyed by the histories
techniques are very important and prove essential for the
construction of a POVM for the time-of-arrival (working however
within the operational formulation of quantum mechanics). The
consideration of measurements smeared in time allows us to
construct a POVM of general validity for the time-of-arrival, in
analogy with POVMs for the probabilities of sequential
measurements. For free particles, this POVM reduces to one
obtained by Kijowski. For a general system, however, the
constructed POVM also depends strongly on the resolution of the
measurement device. This seems to imply that the measurement of
the time-of-arrival is highly contextual within standard quantum
theory.

\section*{Acknowledgments}

C.A. was supported by a Marie Curie Reintegration Grant of the
European Commision, and from  grants by the Empirikion Foundation
and Pythagoras II (EPEAEK), while N.S. was supported by an EPSRC
Research Grant.

\begin{appendix}
\section{The quantum Zeno effect is not robust}

We provide here a simple example demonstrating the  quantum Zeno
effect is not robust, in the sense that even a small deviation
from a projection operator in the definition (\ref{Ct}) of the
operator $\hat{C}_t$ yields to a qualitatively different
behaviour.

We consider a spin system: the Hilbert space is ${\bf C}^2$, and
we choose the Hamiltonian $\hat{H}$ and projector $\hat{E}$ to
correspond to the matrices
\begin{eqnarray}
H = \left( \begin{array}{cc} 0 & \epsilon \\
\epsilon & 0 \end{array} \right) \hspace{2cm} E = \left(
\begin{array}{cc} 0 & 0 \\ 0 & 1 \end{array} \right).
\end{eqnarray}

We consider the self-adjoint operator
\begin{eqnarray}
\hat{V} = \left( \begin{array}{cc} x & 0 \\0 & y \end{array}
\right),
\end{eqnarray}
from which we define the positive operator
\begin{eqnarray}
e^{-V \delta t} = \left( \begin{array}{cc} e^{-x \delta t} & 0\\ 0
& e^{-y \delta t}  \end{array} \right)
\end{eqnarray}
This operator can be seen as a regularised expression for
$\hat{E}$
\begin{eqnarray}
\lim_{x \rightarrow \infty} \lim_{y \rightarrow 0} e^{-V \delta t}
= \hat{E}.
\end{eqnarray}

We may then  write a regularised version $\hat{K}^y_t$ of the
operator $\hat{C}_t = (\hat{E} e^{-i \hat{H}t/n})^n$, such that.
\begin{eqnarray}
\hat{C}_t = \lim_{y \rightarrow 0} \hat{K}^y_t,
 \end{eqnarray}
The operator $\hat{K}^y_t$ reads explicitly
\begin{eqnarray}
\hat{K}^y_t =  \lim_{x \rightarrow \infty} \lim_{n \rightarrow
\infty}(e^{-i\hat{H} t/n} e^{-\hat{V} t/n})^n =  \lim_{x
\rightarrow \infty} e^{-i Ht - Vt}
\end{eqnarray}

We easily find that
\begin{eqnarray}
\hat{K}^y_t = e^{-yt} \hat{E},
\end{eqnarray}
 has the
exponential fall behavior that characterises Fermi's golden rule.

When the limit $y \rightarrow 0$ is also taken, we obtain the
familiar result $\hat{C}_t = \hat{E}$, which is trivially a
degenerate unitary operator. However, even a small deviation from
$\hat{E}$ in the definition of $\hat{C}_t$ leads to a different
(and intuitively more physical) qualitative behaviour.

\end{appendix}

\end{document}